\documentclass[12pt, a4paper, onecolumn, oneside]{article}
\usepackage{amsmath, amsfonts, bbm}
\usepackage{color, bm}
\usepackage{subcaption}
\usepackage{natbib}
\usepackage{graphicx}
\usepackage{multirow}
\usepackage{newtxtext,newtxmath}
\usepackage{geometry}
\usepackage{ulem}
\usepackage[linesnumbered,ruled,vlined]{algorithm2e}
\begin{document}

\title{Evaluating causal effects on time-to-event outcomes in an RCT in oncology with treatment discontinuation}

\author{
Veronica Ballerini (veronica.ballerini@unifi.it) \\
\multicolumn{1}{p{.8\textwidth}}{\centering\textit{Department of Statistics, Computer Science, Applications ``G. Parenti'', University of Florence, Florence, Italy}} \\
Bj\"orn Bornkamp (bjoern.bornkamp@novartis.com) \\
\multicolumn{1}{p{.8\textwidth}}{\centering\textit{Global Drug Development, Novartis Pharma AG, Basel, Switzerland}} \\
Alessandra Mattei (alessandra.mattei@unifi.it)\\
\multicolumn{1}{p{.8\textwidth}}{\centering\textit{Department of Statistics, Computer Science, Applications ``G. Parenti'' \& Florence Center for Data Science, University of Florence, Florence, Italy}} \\
Fabrizia Mealli (Fabrizia.Mealli@eui.eu) \\
\multicolumn{1}{p{.8\textwidth}}{\centering\textit{European University Institute, Fiesole, Italy \& Florence Center for Data Science, University of Florence, Florence, Italy}} \\
Craig Wang (craig.wang@novartis.com)\\
\multicolumn{1}{p{.8\textwidth}}{\centering\textit{Global Drug Development, Novartis Pharmaceuticals Corporation, New Jersey, USA}} \\
Yufen Zhang (yufen.zhang@novartis.com)\\
\multicolumn{1}{p{.8\textwidth}}{\centering\textit{Global Drug Development, Novartis Pharmaceuticals Corporation, New Jersey, USA}} \\
}
\date{}
\maketitle

\abstract
In clinical trials, patients may discontinue treatments prematurely, breaking the initial randomization. 
The ICH E9(R1) Addendum provides guidelines for handling such “intercurrent events;” the right strategy to adopt depends on the questions of interest. 
Stakeholders in drug development are generally interested in going beyond the Intention-To-Treat (ITT) analysis, which provides valid causal estimates of the effect of treatment assignment but does not inform on the effect of the actual treatment receipt. 
Our study is motivated by a randomized controlled trial (RCT) in oncology, where patients assigned the investigational treatment may discontinue it due to adverse events.
We propose adopting a principal stratum strategy and decomposing the overall ITT effect into principal causal effects for groups of patients defined by their potential discontinuation behaviour. 
We first show how to implement a principal stratum strategy to assess causal effects on a survival outcome in the presence of continuous time treatment discontinuation, its advantages, and the conclusions one can draw. 
Our strategy deals with the time-to-event intermediate variable that may not be defined for patients who would not discontinue; moreover, discontinuation time and the primary endpoint are subject to censoring. 
We employ a flexible model-based Bayesian approach to tackle these complexities, providing easily interpretable results. 
We apply this Bayesian principal stratification framework to analyze synthetic data of the motivating oncology trial. We simulate data under different assumptions that reflect real scenarios where patients' behaviour depends on critical baseline covariates. 
Supported by a simulation study, we shed light on the role of covariates in this framework: beyond making structural and parametric assumptions more credible, they lead to more precise inference and can be used to characterize patients' discontinuation behaviour, which could help inform clinical practice and future protocols.\\


\section{Introduction} \label{intro}
Randomized controlled trials (RCTs) are the gold standard for assessing causal effects in medical studies.
RCTs, however, may suffer from complications that are not under experimental control; 
phenomena such as noncompliance, censoring by death, treatment switching, or treatment discontinuation are all events that break initial randomization since they occur after it, and they can either preclude observation of the outcome of interest or affect its interpretation.
The addendum to the E9 guideline on ``Statistical principles in clinical trials'', released by the International Council of Harmonization (ICH), refers to these types of events as \textit{intercurrent events} \citep{ICHE9R1}. 

The case study that motivated this work consists of a clinical trial in which premature treatment discontinuation can occur. Questions were raised to better understand the effect of partial treatment receipt in the subgroup of patients who discontinued treatment. 
In particular, we deal with an RCT in oncology aimed to assess the causal effects of a new investigational drug combined with standard of care versus standard of care only on progression-free survival, i.e., the time from randomization until disease progression or death. 
Patients enrolled in the investigational treatment arm can prematurely discontinue treatment if they experience adverse events (AEs), such as side effects. 
In such a context, and also in line with the tripartite estimand strategy \citep{akacha2017estimands}, relevant questions to patients, physicians, pharmaceutical companies, and regulatory agencies concern i) the treatment effect for patients who adhere to the treatment for its intended duration, ii) the proportion of those who discontinue the investigational treatment prematurely, and iii) the effect for patients who discontinue the treatment prematurely. 
The latter patients could still derive a benefit from taking the treatment initially. 
For example, \cite{schadendorf2017efficacy} investigate such questions from a clinical perspective for a new treatment in oncology.

To face the issues that may concern different stakeholders, we propose to deal with the problem of estimating treatment effects in the presence of premature treatment discontinuation using the principal stratum strategy. This strategy is considered in the ICH E9 (R1) Addendum (\citealp{ICHE9R1}) as one that  provides valid causal estimands in the presence of intercurrent events. 
The principal stratum strategy is based on the principal stratification framework introduced by \citet{frangakis2002principal}. 
Principal stratification (PS) focuses on causal effects for patients belonging to latent subpopulations, namely the ``principal strata,'' defined by the post-randomization variables of interest, in this work, the premature discontinuation behavior.

Treatment discontinuation can be viewed as a form of partial noncompliance, in which patients take only part of the assigned dose.
If we consider the patients who can tolerate the treatment for its intended duration as \textit{compliers}, it is natural to resort to PS. 
Indeed, noncompliance is a natural application of PS since \cite{frangakis2002principal}, and even before their formalization; see \cite{angrist1996identification} and \cite{hirano2000assessing}.
Since then, key methodological contributions to the topic have been made; among many others, see \cite{mealli2004analyzing},
\cite{mattei2007application},
\cite{roy2008principal}, 
\cite{jin2008principal}, \cite{schwartz2011bayesian}, \cite{sheng2019estimating}, \cite{jiang2021identification}, \cite{liu2023principal}, and 
\cite{mattei2024assessing}.

Recent reviews of the principal stratum strategy in the context of clinical trials in drug development were provided by \citet{bornkamp2021principal} and \citet{lipkovich2022using}.
Articles showing the potential of principal stratification often rely on dated case studies \cite[see][among others]{jin2008principal,schwartz2011bayesian}.

Yet, motivated by a recent RCT in oncology, we can deal with and specify a problem-driven model that could address substantive research questions under different scenarios.
In this article, we provide a method to answer questions raised during the last years by the pharmaceutical community \citep{akacha2017estimands, qu2020general, qu2021implementation}, specifically addressing the treatment discontinuation problem in continuous time.
In terms of \textit{efficacy}, we can provide well-defined estimands for measuring treatment effects for those who potentially adhere to the treatment for its intended duration and causal effects for those who potentially discontinue the treatment at a certain time. 
Our model can also estimate the percentage of patients who discontinue treatment and, leveraging the available covariates, the probability that a specific patient will discontinue treatment and after how long. 

Previous works exploited the advantages of a Bayesian approach for inference in a PS framework when dealing with post-randomization events; see \citet{magnusson2019bayesian}, \cite{ohnishi2022bayesian} and \citet{mattei2024assessing}. 
In this work, the Bayesian approach allows us to properly account for the fact that the discontinuation time is either not defined (for those who would never discontinue) or continuous, generating a continuum of principal strata.
Moreover, we consider that both survival time and discontinuation time are subject to censoring.
We use a modelling strategy similar to \cite{mattei2024assessing}, which has proven to be effective in estimating principal causal effects in the presence of treatment switching; however, we target our approach at current policy interests. The general modelling approach proposed by \cite{mattei2024assessing} involves parameters that 
must be fixed a priori and treated as sensitivity parameters; also, it does not allow us to easily test the null hypothesis of no effect. 
Our model gets rid of such parameters, making it much easier and more convenient to assess the effect of a treatment. 
Furthermore, the role of the covariates in such a framework is poorly explored in \cite{mattei2024assessing}. 
Recently, the Food and Drug Administration has released some guidelines on the use of covariates in RCTs, recommending the use of prognostic baseline covariates ``to improve the statistical efficiency for estimating and testing treatment effects.'' \citep[][p. 1]{FDA}. 
However, the utility of covariates goes far beyond the improvement of efficiency.
Supported by a simulation study, we show how the inclusion of baseline covariates, which are good predictors of the principal stratum membership, improves the robustness of the estimates also under misspecification, and it makes the structural and parametric assumptions more credible.
We also show how the covariates can be used to inform about the risk of AEs and characterize the principal strata; this could help in designing future protocols. 
Given the practical use of covariates in our work, we also define principal effects of interest conditioning on their empirical distribution, i.e., what \citep{li2023bayesian} call \textit{mixed} causal effects.

The following Section introduces the case study that motivated this work. 
Section \ref{sec3} presents our principal stratum strategy, recalling the potential outcomes approach and defining the principal strata and the respective principal causal estimands.
Section \ref{sec4} sets the model and the inference in a fully Bayesian framework.
Section \ref{sec5} shows how we perform the analyses under different scenarios.
A Monte Carlo simulation study highlights the role of covariates in improving the precision of causal effects estimates; Section \ref{cov2} shows how covariates can be leveraged to characterize the principal strata.
The discussion follows.
Some details of this work are discussed in the Supplementary Material available online. 

\section{The case study} \label{sec2}
\begin{figure}[t]
    \centering
    \includegraphics[width=.8\linewidth]{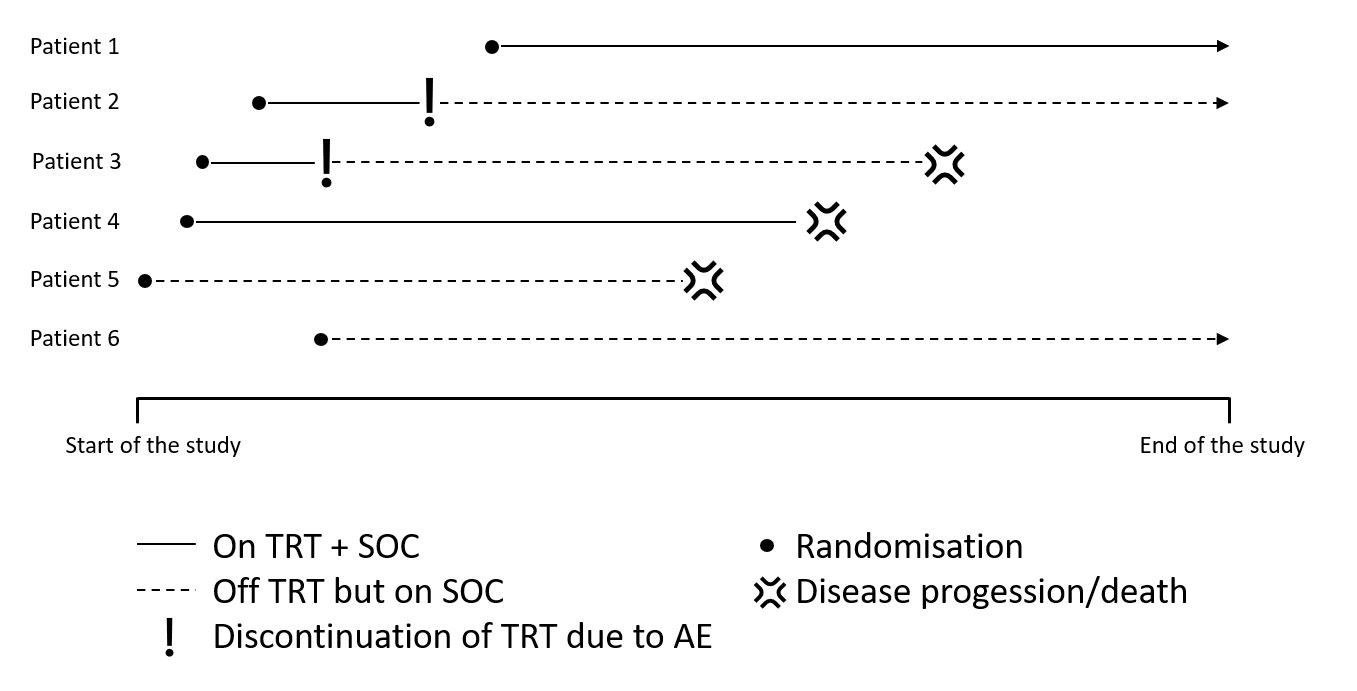}
    \caption{Illustration of patients' journeys in the specific trial. 
    Patients 1 to 4 are assigned to the investigational treatment plus SOC, whereas Patients 5 and 6 are assigned to the SOC only. 
    During the follow-up period, in the treatment arm: Patient 1 remains on treatment for its intended duration; Patient 2 discontinues the treatment due to AE and continues with SOC; Patient 3 discontinues due to AE and then experiences a primary event; Patient 4 experiences the primary event without discontinuing the treatment.
    On the control arm: Patient 5 experiences the primary event, while Patient 6 does not.}
    \label{fig:paths}
\end{figure}

The case study that motivated this work deals with a recent RCT in oncology, aiming to assess the causal effects of a new investigational drug combined with the standard of care (SOC) versus SOC on progression-free survival, i.e., the time from randomization until a primary event that can be either disease progression or death. 
The study duration from the first randomized patient to the analysis cutoff date is approximately $c = 33$ months.
All patients were enrolled during the first 23 months; 
thus, each patient can have a different follow-up period (or time to censoring) $C_i \in [10, 33]$. 
Let $Z_i$ denote the treatment assigned to the $i^{th}$ patient, and let $z$ denote a realization of $Z_i$; $z$ can be either $1$ if $i$ is assigned to the new investigational treatment - in addition to SOC (investigational treatment), or $0$ if $i$ is assigned SOC only (control treatment).
Among $n=335$ patients, $n_1 = 181$ were assigned to the new investigational treatment, and $n_0 = 154$ were assigned to SOC.
When patients in the new treatment arm incurred AEs, they were allowed to discontinue the new investigational treatment but continued on SOC.
On the contrary, patients under control treatment could not receive the investigational treatment.
Figure \ref{fig:paths} shows the possible patients' journeys.

Let $D_i$,  $Y_i$, and $C_i$ denote the true underlying time to discontinuation, progression-free survival time and censoring time, respectively. 
We denote with $\tilde{D}_i = \min\{D_{i}, C_i\}$ and $\tilde{Y}_i = \min\{Y_{i}, C_i\}$ the censored discontinuation time and the time to primary event, respectively.
The baseline information available consists of the following covariates:
\begin{itemize}
    \item $X_1$: continuous variable. 
    The higher the value of $X_1$, the higher the risk of progression;
    \item $X_2$: binary indicator for metastatic status. $X_2 = 1$ denotes higher progression risk;
    \item $X_3$: binary indicator for disease burden. As for $X_2$, $X_3 = 1$ denotes higher progression risk.
\end{itemize}
For confidentiality reasons, here we use summary statistics based on synthetic data generated using real case data.
The statistics are summarised in Tables \ref{novartis_synth} and \ref{tab:my-table}. 
This gives us the opportunity to simulate individual data under different hypothetical assumptions of treatment effects to test our model performance; see Section \ref{sec5}.

\begin{table}
\centering
\caption{Summary statistics of synthetic data. Here $n=335$, $n_1 = 181$, and $n=154$.} 
\label{novartis_synth}
{\begin{tabular}{@{}llcccccc@{}}
\hline
\textbf{Variable}    & \textbf{Mean(proportion)} & \textbf{SD} & \textbf{Min} & \textbf{Q1} & \textbf{Median} 
& \textbf{Q3} & \textbf{Max} \\ \hline
$Z = 1$                & 54.00\% (181/335)          & -----       & -----        & -----       & ----- 
& -----       & -----  \\ 
$\mathbb{I}(D<C)$   & 16.57\%(30/181)  & -----       & -----        & -----  & -----        
& ----- & ----- \\
$\tilde{D}$     & 3.72            & 5.04         & 0.26       & 1.07   & 1.41 & 
4.09 & 24.11\\ 
$\mathbb{I}(Y<C)$ & 66.00\% (221/335)    & ----- & ----- & ----- & ----- 
& ----- & -----\\
$\tilde{Y}$        & 6.60  & 5.95 & 0.10 & 1.87 & 3.81 
& 9.17 & 27.63 \\  \textbf{Continuous covariates}                  &                  &        &       &         \\ 
 $X_1$                                     & 63.27           & 10.50 & 25.00  & 57.00 & 63.00 & 71.00 & 92.00  \\ 
 \textbf{Categorical covariates}    &      &  &  &  &  
&  & \\
 $X_2$ & 43.90\% (147/335)    & ----- & ----- & ----- & ----- 
& ----- & -----\\
$X_3$ & 23.90\% (80/335)    & ----- & ----- & ----- & ----- 
& ----- & -----\\
\hline
\end{tabular}}
\end{table}

\begin{table}[]
\centering
\caption{Summary statistics of synthetic data, detail. Incidence of primary and adverse events during the follow-up by treatment status.}
\label{tab:my-table}
\begin{tabular}{lccccccccc}
                                      &                        & \multicolumn{2}{c}{Z=1}                            &                      &                      &                      & \multicolumn{2}{c}{Z = 0}                &                      \\ \cline{3-4} \cline{8-9}
                                      & \multicolumn{1}{l}{}   & \multicolumn{2}{l}{$\mathbb{I}(Y\le C)$}           & \multicolumn{1}{l}{} & \multicolumn{1}{l}{} & \multicolumn{1}{l}{} & \multicolumn{2}{l}{$\mathbb{I}(Y\le C)$} & \multicolumn{1}{l}{} \\ \cline{3-4} \cline{8-9}
                                      &                        & \multicolumn{1}{c|}{1}   & 0                       &                      &                      &                      & \multicolumn{1}{c|}{1}         & 0       &                      \\ \cline{2-5} \cline{8-9}
\multirow{2}{*}{$\mathbb{I}(D\le C)$} & \multicolumn{1}{c|}{1} & \multicolumn{1}{c|}{21}  & \multicolumn{1}{c|}{9}  & 30                   &                      &                      & \multicolumn{1}{c|}{110}       & 44      & 154                  \\ \cline{2-5} \cline{8-10} 
                                      & \multicolumn{1}{c|}{0} & \multicolumn{1}{c|}{90}  & \multicolumn{1}{c|}{61} & 151                  &                      &                      &                                &         &                      \\ \cline{2-5}
                                      &                        & \multicolumn{1}{c|}{111} & \multicolumn{1}{c|}{70} & 181                  &                      &                      &                                &         &                      \\ \cline{3-5}
\end{tabular}
\end{table}

\section{Methods. A principal stratum strategy}
\label{sec3}
Consider the randomized controlled clinical trial described in the previous Section.
Applying the widespread Intention-To-Treat (ITT) principle would provide valid causal estimates of the treatment assignment, neglecting the premature treatment discontinuation;
as the name suggests, it is a treatment policy estimand and does not always inform on treatment efficacy.

We apply the principal stratification framework \citep{frangakis2002principal} and define causal estimand on the treatment effect that can capture the effect heterogeneity with respect to the discontinuation time.

\subsection{Potential outcomes approach} \label{po}
Principal stratification heavily relies on the potential outcomes (or ``Rubin causal'') model \citep{rubin1974estimating}.
Here, we describe the structural assumptions we do and the potential outcomes we need to build our model.

Let $Y_i(z)$ be the potential progression-free survival times for the $i^{th}$ unit under treatment assignment $z$. 

\paragraph{Assumption 1} \textit{Stable Unit Treatment Value Assumption} \cite[SUTVA;][]{rubin1980randomization}, which implies
\begin{equation}
    P(Z_i\mid Y(1), Y(0), D_i(1), X) = P(Z_i \mid Y_i(1), Y_i(0), D_i(1), X_i)\;, \quad  \forall i = 1, \dots, n
\end{equation}
\begin{equation}
    Y_i = Z_i Y_i(1) + (1-Z_i) Y_i(0)\;, \quad  \forall i = 1, \dots, n
\end{equation}

Let $D_i(z)$ be the potential time to discontinuation of unit $i$ under treatment assignment $z$.
Since the discontinuation of the new treatment is possible only under investigational treatment, the discontinuation time is not defined under control.
This scenario is similar to a randomized study with one-sided partial compliance, although the time-to-event nature of the discontinuation behavior makes it particularly challenging.
Following \cite{mattei2024assessing}, we formally set $D_i(0) = \bar{\mathbb{D}} ~\forall i$, meaning that $D_i(0)$ takes on a non-real value.
Yet, patients are allowed to discontinue the investigational treatment as long as they do not experience the primary event. 
We say that $D_i(1)$ is censored by death or progression; indeed, $D_i(1) < Y_i(1)$.
Therefore, on the one hand, for those who potentially discontinue, $D_i(1)$ is defined in $\mathbb{R}^+$. 
On the other hand, for the units who experience the event without discontinuing (either before or after the follow-up period), the time to discontinuation is not defined, i.e., $D_i(1) = \bar{\mathbb{D}}$. 

We assume initial randomization, which holds by design in an RCT:

\paragraph{Assumption 2} \textit{Randomization}
\begin{equation}\label{randomization}
P(Z_i \mid Y_i(1), Y_i(0), D_i(1), X_i) = P(Z_i) \;, \quad \forall i = 1, \dots, n
\end{equation}

The study duration is limited to $c$ months, and each patient can enter the study at a different time, implying different censoring times $C_i(z) \leq c$ under treatment assignment $z$ for different patients.
Since the censoring is administrative, it is sensible to assume:

\paragraph{Assumption 3} \textit{No effect of the treatment assignment on the censoring} 
\begin{equation}
C_i(1) = C_i(0) = C_i\;, \quad \forall i = 1, \dots, n
\end{equation} 

\paragraph{Assumption 4} \textit{Ignorable censoring}
\begin{equation}\label{ign_cens}
P(C_i\mid Y_i(1),Y_i(0),D_i(1)) = P(C_i) \;, \quad \forall i = 1, \dots, n
\end{equation}

\subsection{Principal strata and causal estimands} \label{ps_estimands}
According to their discontinuation behavior, the units are classified into different (latent) subpopulations, or \textit{principal strata}.
Since $D_i(0) = \bar{\mathbb{D}}$, $\forall i$, 
treatment discontinuation is one-sided; thus, the strata are defined exclusively with respect to the discontinuation behavior under the new investigational treatment, $D_i(1)$.
Since $D_i(1)$ is either not defined or continuous, units can be either patients that would not discontinue if assigned to the investigational treatment, i.e., those $i$ such that $D_i(1) = \bar{\mathbb{D}}$, or patients that would discontinue at a time $d \in \mathbb{R}^+$.
Throughout the paper, we refer to the former patients as ``Never Discontinuing'' (ND patients) and to the latter as ``Discontinuing'' (D patients).
Given that the discontinuation variable is continuous, the group of patients who would discontinue at some point in time is the union of infinite basic principal strata D.

Under such a principal stratification approach, we propose to decompose the ITT effect 
\begin{equation}\label{itteffect}
    ITT = \mathbb{E}(Y_i(1) - Y_i(0))
\end{equation}
defining \textit{principal causal estimands}, namely causal estimands for each latent subpopulation.
Characterizing the ITT effect heterogeneity with respect to the discontinuation time is challenging because the discontinuation can be \textit{non-ignorable}, i.e., the discontinuation status is not necessarily independent of the potential outcomes even conditional on covariates: $Y_i(1), Y_i(0) \not\!\perp\!\!\!\perp  D_i(1) | X_i$.

In the case of \textit{average} causal effects, the ITT effect would be a weighted average of \textit{principal} average causal effects, i.e., average causal effects conditional for the discontinuation behaviour:
\begin{equation}\label{itt}
      ITT = \pi_{\text{ND}}ACE_\text{ND} + (1-\pi_{\text{ND}})ACE_\text{D}\;,
\end{equation}
where $\pi_{\text{ND}}$ is the proportion of ND patients, 
and $ACE_\text{ND}$ and $ACE_{\text{D}}$ are the average causal effects for ND and D patients, respectively. 

First, let us define the conditional $ACE_{\text{ND}}$ and $ACE_{\text{D}}$ on covariates value $x$:
\begin{equation}
     ACE_\text{ND}(x) \equiv \mathbb{E}\left[Y_i(1)\mid D_i(1) = \bar{\mathbb{D}}, X_i = x\right]-\mathbb{E}\left[Y_i(0)\mid D_i(1) = \bar{\mathbb{D}}, X_i= x\right]
\end{equation}
\begin{equation}
     ACE_\text{D}(d,x) \equiv \mathbb{E}\left[Y_i(1)\mid D_i(1) = d, X_i = x\right]-\mathbb{E}\left[Y_i(0)\mid D_i(1) = d, X_i= x\right] \quad \forall d\in \mathbb{R}^+ \; .
\end{equation}
We may define \textit{mixed} principal average causal effects, namely conditional principal causal effects averaged over the empirical distribution of the covariates \citep{li2023bayesian}, as follows:
\begin{itemize}
    \item 
For ND patients, we write the mixed principal average causal effect as
\begin{equation}\label{pace_nd}
\begin{split} 
    ACE_\text{ND} & = \sum\limits_{i=1}^{N}{ACE_\text{ND}(x_i)P(X_i=x_i|D_i(1)=\bar{\mathbb{D}})} = \\ 
    & =  \sum\limits_{i=1}^{N}{ACE_\text{ND}(x_i)\dfrac{P(D_i(1)=\bar{\mathbb{D}}\mid x_i)P(X_i = x_i)}{\sum P(D_i(1)=\bar{\mathbb{D}}\mid x_i)P(X_i = x_i)}} = \\ 
    & =  \sum\limits_{i=1}^{N}{ACE_\text{ND}(x_i)\dfrac{P(D_i(1)=\bar{\mathbb{D}}\mid x_i)}{\sum P(D_i(1)=\bar{\mathbb{D}}\mid x_i)}} 
    \end{split}
\end{equation}
where the last equality follows from setting $P(X_i = x_i) = 1/N$. 

\item For D patients:
\begin{equation}
      ACE_\text{D} = \int_{\mathbb{R}^+}{ACE_{\text{D}}(d)f_D(d)}\;\mathrm{d}d \; ,
\end{equation}
where $f_D(d)$ is the density of $D(1)$, and similarly to \eqref{pace_nd}
\begin{equation}\label{pace_d}
\begin{split}
    ACE_\text{D}(d) & = \sum\limits_{i=1}^{N}{ACE_\text{D}(d,x_i)\dfrac{P(D_i(1)\in \mathbb{R}^+ \mid x_i)}{\sum P(D_i(1) \in \mathbb{R}^+ \mid x_i)}} \;.
    \end{split}
\end{equation} 
\end{itemize}

When dealing with survival outcomes, one may be interested in the \textit{mixed distributional causal effects of the treatment} (or \textit{mixed principal survival difference}) at time $y$. 
The mixed principal distributional causal effect at $y$ for the set of ND patients is:
\begin{equation}\label{dce_nd}
\begin{split}
        DCE_{\text{ND}}(y) =  \sum\limits_{i=1}^{N}DCE_{\text{ND}}(y\mid x_i) \dfrac{P(D_i(1)=\bar{\mathbb{D}}\mid x_i)}{\sum P(D_i(1)=\bar{\mathbb{D}}\mid x_i)} \;, \quad \forall y \in \mathbb{R}^+\;,
\end{split}
\end{equation}
where 
\begin{equation}
   DCE_{\text{ND}}(y \mid x_i) = P(Y_i(1)>y \mid D_i(1) = \bar{\mathbb{D}}, X_i = x_i) - P(Y_i(0)>y\mid D_i(1) = \bar{\mathbb{D}}, X_i= x_i)\;,
\end{equation}
whereas the mixed distributional causal effects of the treatment at time $y$ for D patients discontinuing in $d$ would be
\begin{equation}\label{dce_d}
\begin{split}
        DCE_\text{D}(y\mid d)  = \sum\limits_{i=1}^{N} DCE_{\text{D}}(y\mid d, x_i) \dfrac{P(D_i(1)\in \mathbb{R}^+ \mid x_i)}{\sum P(D_i(1)\in \mathbb{R}^+ \mid x_i)}, \quad y \in \mathbb{R}_+ \;.
    \end{split}
\end{equation}
where
\begin{equation}
    DCE_{\text{D}}(y\mid d, x_i)  = P(Y_i(1)>y \mid D_i(1) = d , X_i = x_i) - P(Y_i(0)>y\mid D_i(1) = d, X_i= x_i)\;.
\end{equation}

\subsection{Potential outcomes and observed data} \label{po_obsdata}
Considering the censoring, once the treatment has been assigned, the observed progression-free survival and the observed discontinuation time for each unit $i$ are
\begin{equation}
   \tilde{Y}_i = \min\{Y_i, C_i\} = \min\{Z_i Y_i(1) + (1-Z_i)Y_i(0),C_i\}
\end{equation}
 and
\begin{equation}
\tilde{D}_i = 
    \begin{cases}
    \min{\{D_i,C_i\}} = \min\{Z_iD_i(1) + (1-Z_i)\bar{\mathbb{D}},C_i\} & \forall i:Z_i = 1,D_i \in \mathbb{R}^+
    \\
    C_i  & \forall i:D_i = \bar{\mathbb{D}}  
    \end{cases}
\end{equation}
According to their behaviour, it is possible to categorize different patients' profiles. 
Table \ref{tab_t} summarises the observed profile.
Under treatment, among those who have experienced the primary event during the follow-up ($\tilde{Y}_i = Y_i$), there may be either D patients, namely, those who had discontinued the treatment ($\tilde{D}_i = D_i$; Patient 3, Figure \ref{fig:paths}), and patients who had not ($\tilde{D}_i = C_i$; Patient 4, Figure \ref{fig:paths}) and hence are ND patients undoubtedly.
Among those who have not experienced the primary event during the follow-up ($\tilde{Y}_i = C_i$), there may be either patients who have discontinued the treatment (D patients; Patient 2, Figure \ref{fig:paths}) and patients who have not (Patient 1, Figure \ref{fig:paths}); the latter set is a mixture of ND patients and D patients at time $d > C_i$.
Under control, we cannot observe the patients' behavior under treatment; their profiles are thus infinite mixtures of D patients at time $d,  0 < d < Y_i$, and ND patients.

\begin{table}[]
\centering
\caption{Observed profiles according to the observed discontinuation behavior.} 
\label{tab_t}
{\begin{tabular}{@{}lllc@{}}
$Z_i$ & $\tilde{Y}_i$ & $\tilde{D}_i$ & Principal stratum label \\ \hline
1     & $Y_i$         & $D_i$        & D                       \\
1     & $Y_i$         & $C_i$                     & ND                      \\
1     & $C_i$                     & $D_i$        & D                       \\
1     & $C_i$                     & $C_i$                     & D or ND                 \\
0     & $Y_i$         & $C_i$                     & D or ND                 \\
0     & $C_i$                     & $C_i$                     & D or ND                 \\ \hline
\end{tabular}}
\end{table}

\section{Bayesian inference}\label{sec4}
Even under initial randomization and the assumption of completely ignorable censoring (Assumptions 2 and 4), the causal estimands described in Section \ref{ps_estimands} are not fully nonparametrically identifiable from the observed data. 

A Bayesian approach would not require full identification \citep{lindley1972bayesian}; in the case of weak or partial identifiability, i.e., when the posterior shows regions of flatness, the related uncertainty would be properly quantified \citep{gustafson2009limits}.
Hence, aiming to estimate the causal effect of the treatment for each stratum, we adopt a parametric Bayesian approach. 

In the following section, we provide a general setting to model the intermediate variable and the potential outcomes given the intermediate variable.

\subsection{Model setting}
To each unit $i$ we associate the following quantities: $Z_i, C_i, Y_i(1), Y_i(0), D_i(1), X_i$.
We observe $Z_i, C_i, \tilde{Y}_i, \tilde{D}_i, X_i$ for each $i$. 
Indeed, $Y_i(Z_i)$ is only observed for units that experienced the event under treatment $Z_i$; $D_i(1)$ is only observed for some $i$, namely the D patients assigned to treatment whose discontinuation time is not censored.
However, $Y_i(1-Z_i)$ is missing for all $i$.
Note that, under a Bayesian approach, $Y_i$ is a realization of $Y_i(1)$ for the units assigned to treatment and of $Y_i(0)$ for those assigned to the control. 
Similarly, $D_i$ is a realisation of $D_i(1)$ for the units assigned to treatment; yet, $D_i$ is deterministically equal to $\bar{\mathbb{D}}$ for those assigned to control.

Let the joint distribution of the quantities described above be $\pi(Z_i, C_i, Y_i(1), Y_i(0), D_i(1), X_i)$. 
Under exchangeability, 
\begin{equation}\label{joint}
\pi(Z, C, Y(1), Y(0), D(1), X) =   
\int_\Theta \prod_{i=1}^{n} \pi (Z_i, C_i, Y_i(1), Y_i(0), D_i(1), X_i\mid  \theta)
			\pi(\theta) \mathrm{d} \theta
\end{equation}
with $\theta$ being the parameter (vector) that governs the joint distribution. 
Equation \eqref{joint} is equal to
\begin{equation}
\begin{split}
      \int_\Theta \prod_{i=1}^{n} & 	\pi(Z_i \mid Y_i(1), Y_i(0), D_i(1), C_i, X_i; \theta) \pi( C_i \mid Y_i(1), Y_i(0), D_i(1),  X_i;\theta) \\ & 
			\pi(Y_i(0), Y_i(1) \mid D_i(1), X_i; \theta) 
   \pi (D_i(1) \mid  X_i; \theta )  
    \pi(X_i \mid \theta ) 
		\pi(\theta) \mathrm{d} \theta 
\end{split}
\end{equation}

Under \eqref{randomization} and \eqref{ign_cens}, we can write
\begin{equation}
\begin{split}
    \pi(Z_i, C_i, Y_i(1), Y_i(0), D_i(1), X_i) \propto  \int_\Theta \prod_{i=1}^{n} 
    \pi(Y_i(0),~& Y_i(1) \mid D_i(1), X_i; \theta) \\
    & \pi (D_i(1) \mid  X_i; \theta ) \pi(X_i \mid \theta ) 
	\pi(\theta) \mathrm{d} \theta \; .
\end{split}
\end{equation}
We only need to model $(Y_i(0), Y_i(1) \mid D_i(1), X_i)$ and $(D_i(1) \mid  X_i; \theta )$ to make inference on the causal estimands described in Section \ref{ps_estimands}. 
No additional structural assumptions, such as \textit{principal ignorability} or exclusion restriction, are made \citep{mattei2023assessing}.

Let us start defining a two-part model for the discontinuation variable;
the first categorical part models the membership to the group of ND patients versus the group of D ones.
We denote with
\begin{equation} \label{disc_disc}
    \text{I}^{\text{ND}}_i = \mathbb{I}_{\{D_i(1) = \mathbb{D}\}} = \text{I}^{\text{ND}}_i(\theta^{\bar{\mathbb{D}}}, X_i) \; ,
\end{equation}
the discontinuation indicator taking value $1$ when the unit $i$ is ND; we let the probability of $\text{I}^{\text{ND}}_i$ be function of the $3$-dimensional vector of covariates $X_i = (X_{i,1}, X_{i,2}, X_{i,3})$ and of parameters $\theta^{\bar{{\mathbb{D}}}}$.

The second part of the model deals with the potential time-to-discontinuation $D_i(1)$, conditional on the discontinuation status and the covariates $X_i$.
We assume
\begin{equation} \label{cont_disc}
    D_i(1)|\text{I}^{\text{ND}}_i,X_i 
    \begin{cases}
    = \bar{\mathbb{D}} & \text{if } \text{I}^{\text{ND}}_i = 1\\
    \sim \psi_D(\cdot ; \theta^D, X_i) & \text{if } \text{I}^{\text{ND}}_i = 0 \; ,
    \end{cases}
\end{equation}
i.e., the discontinuation time is not defined for the ND patients, while for D patients it follows a generic suitable distribution $\psi(\cdot; \theta^D, X_i)$ depending on some parameters $\theta^{D}$ and covariates $X_i$.

Then, we model the potential survival outcomes assuming that they are conditionally independent given the discontinuation status and time and given the covariates and the parameters' vector. 

Let the potential outcome under treatment conditionally be:
\begin{equation}
    Y_i(1)|\text{I}^{\text{ND}}_i, D_i(1),X_i \begin{cases}
    \sim \psi_{\bar{1}}(\cdot; \bar{\theta}^{1}, X_i) & \text{if }  \text{I}^{\text{ND}}_i = 1\\
    \sim \psi_{1}(\cdot; \theta^{1}, X_i, D_i(1)) & \text{if } \text{I}^{\text{ND}}_i = 0 \; .
    \end{cases}
\end{equation}
By the natural constraint, $Y_i(1) > D_i(1)$; thus, the conditional potential outcome under investigational treatment for D patients must be a truncated variable.

We assume the potential outcome under control $Y_i(0)$ to be independent of $Y_i(1)$ given $D_i(1)$ and the covariates but not independent of the discontinuation status:
\begin{equation}
    Y_i(0)|\text{I}^{\text{ND}}_i, D_i(1),X_i \begin{cases}
    \sim \psi_{\bar{0}} (\cdot;\bar{\theta}^0, X_i) & \text{if }  \text{I}^{\text{ND}}_i = 1\\
    \sim \psi_0(\cdot; \theta^0, X_i, D_i(1)) & \text{if } \text{I}^{\text{ND}}_i = 0 \; .
    \end{cases}
\end{equation}
Here, the potential outcome under control for D patients can also be a function of the discontinuation time. 
Since we never observe $Y_i(0)$ and $D_i(1)$ jointly, the parametric assumptions we make are crucial. 
In Section \ref{model_details}, we suggest a way to model such a relation by exploiting the information given by that of $Y_i(1), D_i(1)$.

We assume that the elements of the parameter vector $\theta = (\theta^{\bar{\mathbb{D}}},\theta^D,\bar{\theta}^1,\theta^1,\bar{\theta}^0,\theta^0)$ are a priori independent.
Hence, we write the joint prior distribution of $\theta$ as
\begin{equation}
    \pi(\theta) = \pi(\theta^{\bar{\mathbb{D}}})\pi(\theta^D)\pi(\bar{\theta}^1)\pi(\theta^1)\pi(\bar{\theta}^0)\pi(\theta^0) \; .
\end{equation}

\subsection{Posterior computation}\label{sec:postcomp}
We aim to draw from the posterior distribution
\begin{equation}\label{posterior}
    \pi(\theta\mid X, {Z},{C},\tilde{{D}},\tilde{{Y}}) \propto \mathcal{L}(\tilde{{Y}},\tilde{{D}}; \theta,X,{Z},{C})\pi(\theta)  \; ,
\end{equation}
where $\mathcal{L}(\tilde{{Y}},\tilde{{D}}; \theta,X,{Z},{C})$ is the observed data likelihood. 
Given the presence of infinite mixtures in the likelihood function, following \cite{mattei2024assessing}, we rely on a data augmentation procedure and estimate the posterior via MCMC.
The computational details can be found in the Supplementary Material.
Note that once we can draw from \eqref{posterior}, we can estimate the posterior distribution of any valid causal estimand of interest beyond those introduced in Section \ref{ps_estimands}.

\section{Application and results}
\label{sec5}
\subsection{Synthetic data}\label{synthetic}
In this article, we illustrate two scenarios of interest.
The first scenario depicts a situation in which the principal causal effects are positive in all the latent strata, reflecting the efficacy of the treatment.
The second scenario represents a more challenging case of a positive overall effect, i.e., $ITT > 0$, but the treatment assignment has no effect for D patients; $ACE_{\text{D}} = 0$. 
In other words, we mimic a situation in which the treatment does not show efficacy due to discontinuation.
The summary statistics of the data simulated under such scenarios are very close between them and similar to the summaries of the real data. 
In the Supplementary Material, we describe in detail the data generating process, and we show the summary statistics for a data set generated under each scenario and used to produce results in Section \ref{sec:scenarios}. 

In both scenarios, we make some assumptions on discontinuation for synthetic data simulation.
\begin{itemize}
    \item The higher the value of $X_1$, the lower the probability of being an ND patient.
    \item Patients with $X_2$, $X_3$ equal to $1$ are more likely to be ND patients, i.e., patients who experience progression-free survival (PFS) without discontinuing.
    \item For D patients, higher-risk patients are more likely to discontinue sooner.
\end{itemize}
For convenience, we standardise the continuous covariate $X_1$, and we use the standardised version in the data-generating process and in the estimation. 
With a little abuse of notation, we continue denoting $X$ the vector of covariates including the standardised $X_1$.

\subsubsection{Scenario I: Positive causal effect for ND and D patients}

We generate data such that there is a positive treatment effect for all latent strata and, thus, a positive effect of the treatment for the ND patients.
To simulate such a situation, we first draw 
I$^{\text{ND}}_i$ from a Bernoulli($p(X_i)$), where
\begin{equation}\label{psmemberI}
    p(X_i) = \dfrac{\exp(\gamma_0 + X_i'{\gamma})}{1+\exp(\gamma_0 + X_i'{\gamma})} \; ;
\end{equation}
then, we specify the model for the potential outcomes as follows:
\begin{equation}\label{timetoDI}
    D_i(1) \mid \text{I}^{\text{ND}}_i = 0 \sim \text{Weibull}(\alpha_D, e^{-(\beta_D+X'_i{\eta}_D)/\alpha_D})
\end{equation}
\begin{equation} \label{sc1_nd1}
     Y_i(1) \mid \text{I}^{\text{ND}}_i = 1, X_i \sim \text{Weibull}(\bar{\alpha}_1, e^{-(\bar{\beta}_1 + X_i'\bar{{\eta}}_1)/\bar{\alpha}_1} )
\end{equation}
\begin{equation} \label{sc1_nd0}
     Y_i(0) \mid \text{I}^{\text{ND}}_i = 1, X_i \sim \text{Weibull}(\bar{\alpha}_0, e^{-(\bar{\beta}_0 + X_i'\bar{{\eta}}_0)/\bar{\alpha}_0} )
\end{equation}
\begin{equation} \label{sc1_d1}
     Y_i(1)\mid \text{I}^{\text{ND}}_i = 0, D(1), X_i \sim \text{tWeibull}_{D_i(1)}(\alpha_1, e^{-(\beta_1 + X_i'{\eta}_1 + \delta\log(D_i(1)))/\alpha_1})
\end{equation}
\begin{equation} \label{sc1_d0}    
     Y_i(0)\mid \text{I}^{\text{ND}}_i = 0, D(1), X_i \sim \text{Weibull}(\alpha_0, e^{-(\beta_0 + X_i'{\eta}_0+ \delta\log(D_i(1)))/\alpha_0})
\end{equation}
In Equation \eqref{sc1_d1}, tWeibull$_{D_i(1)}$ stands for left truncated Weibull with truncation parameter $D_i(1)$. 

\begin{figure}[t]
\centering
\begin{subfigure}{.5\linewidth}
  \centering
  \includegraphics[width=\linewidth]{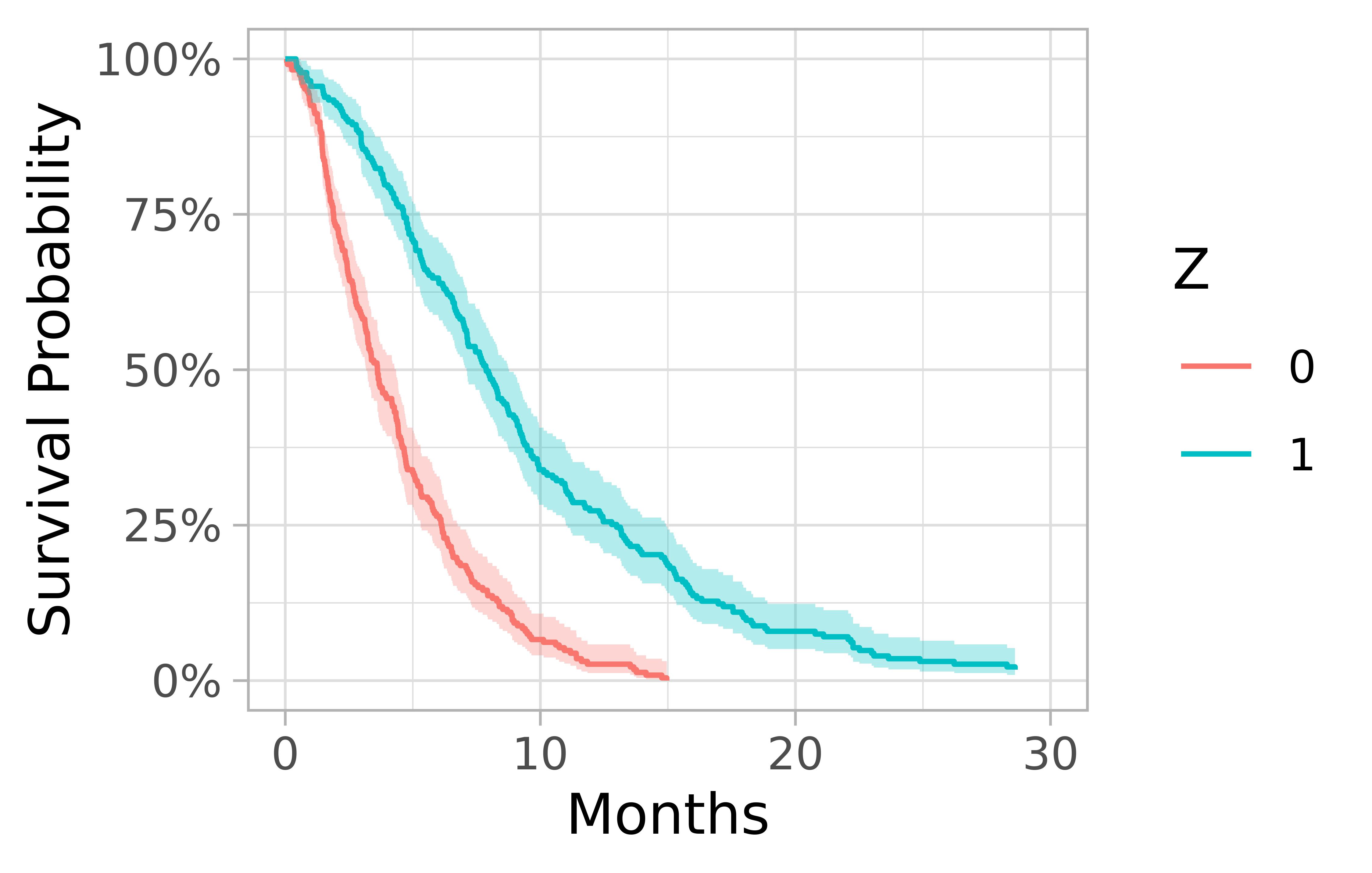}
  \caption{I$^{\text{ND}} = 1$}
  \label{kapmei1_ND}
\end{subfigure}%
\begin{subfigure}{.5\linewidth}
  \centering
  \includegraphics[width=\linewidth]{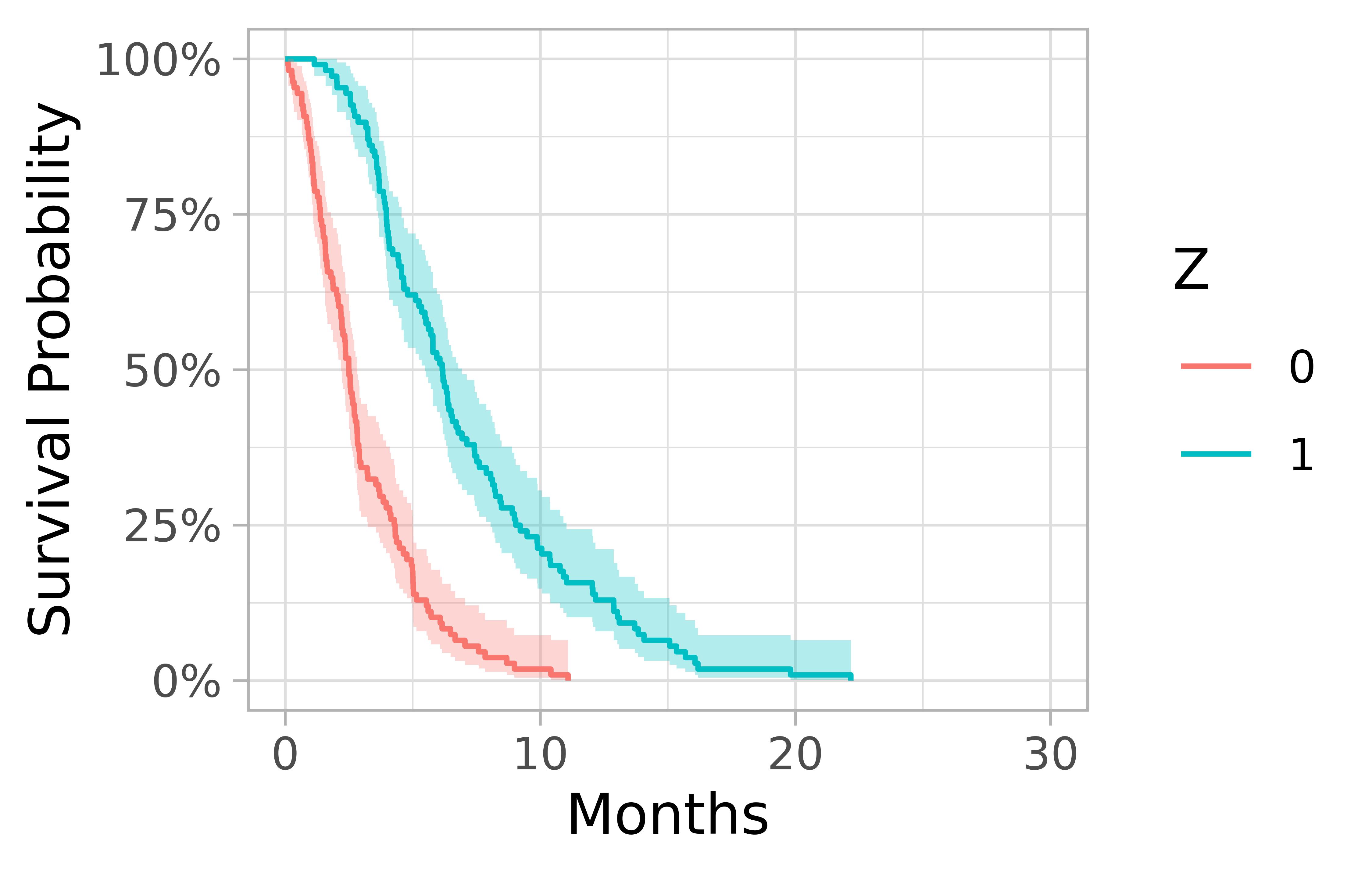}
  \caption{I$^{\text{ND}} = 0$}
  \label{kapmei1_D}
\end{subfigure}
\caption{\textit{Scenario I}. Kaplan-Meier curves for treated (blue) and controls (red), estimated using the potential outcomes' complete simulated data (one sample).}
\label{kapmei1_ID}
\end{figure}

\begin{figure}[t]
    \centering
    \includegraphics[width=0.5\linewidth]{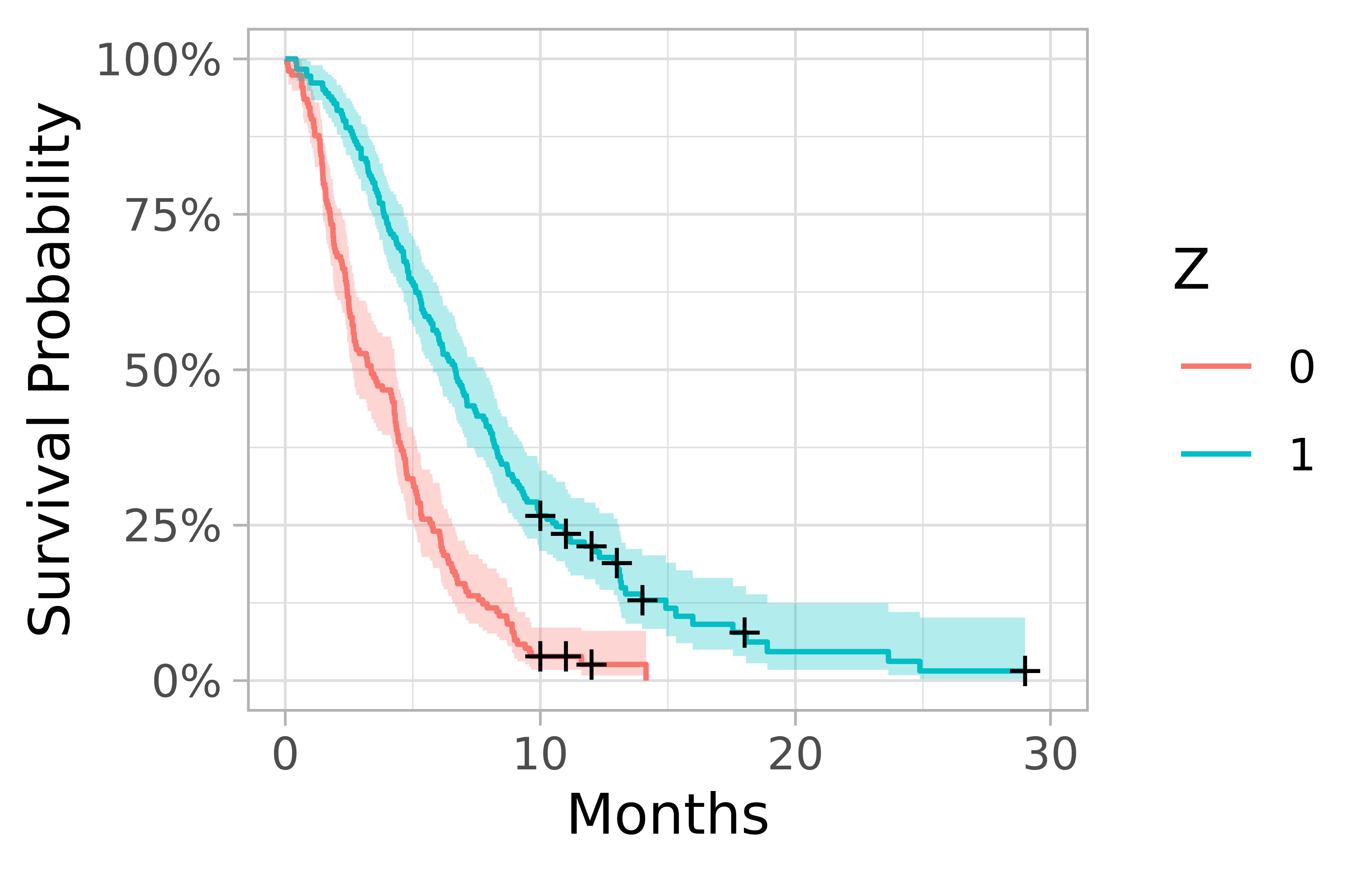}
    \caption{\textit{Scenario I}. Kaplan-Meier curves for treated (blue) and controls (red), estimated using the simulated data (one sample).}
    \label{kapmei1}
\end{figure}

The Kaplan-Meier estimated using the potential outcomes' complete data are shown in Figures \ref{kapmei1_ND} and \ref{kapmei1_D}, whereas Figure \ref{kapmei1} shows the observed survival curves.

\subsubsection{Scenario II: Positive causal effect for ND patients, zero effect for D patients}

\begin{figure}[t]
\centering
\begin{subfigure}{.5\linewidth}
  \centering
  \includegraphics[width=\linewidth]{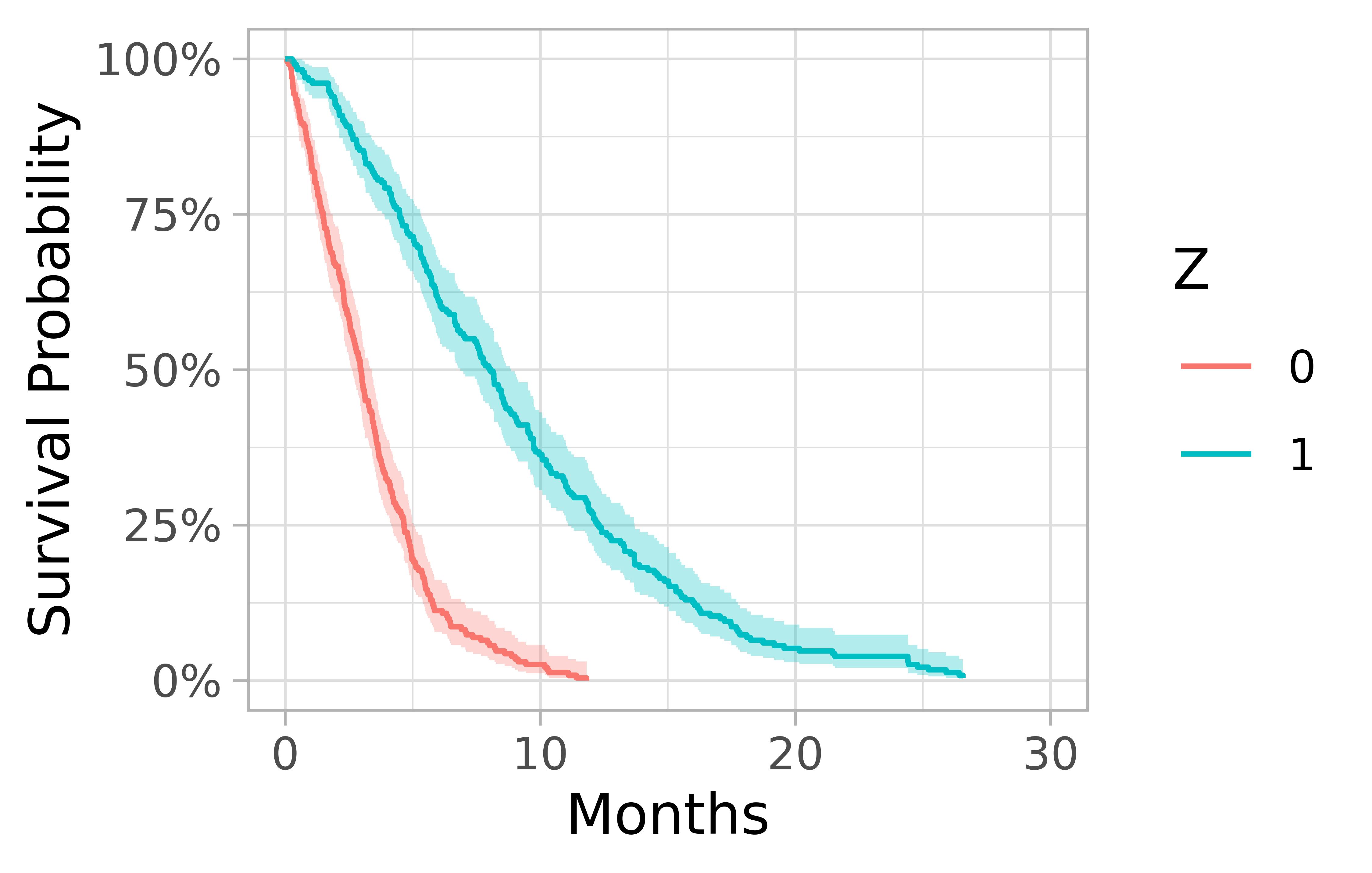} 
  \caption{I$^{\text{ND}} = 1$}
  \label{kapmei2_ND}
\end{subfigure}%
\begin{subfigure}{.5\linewidth}
  \centering
  \includegraphics[width=\linewidth]{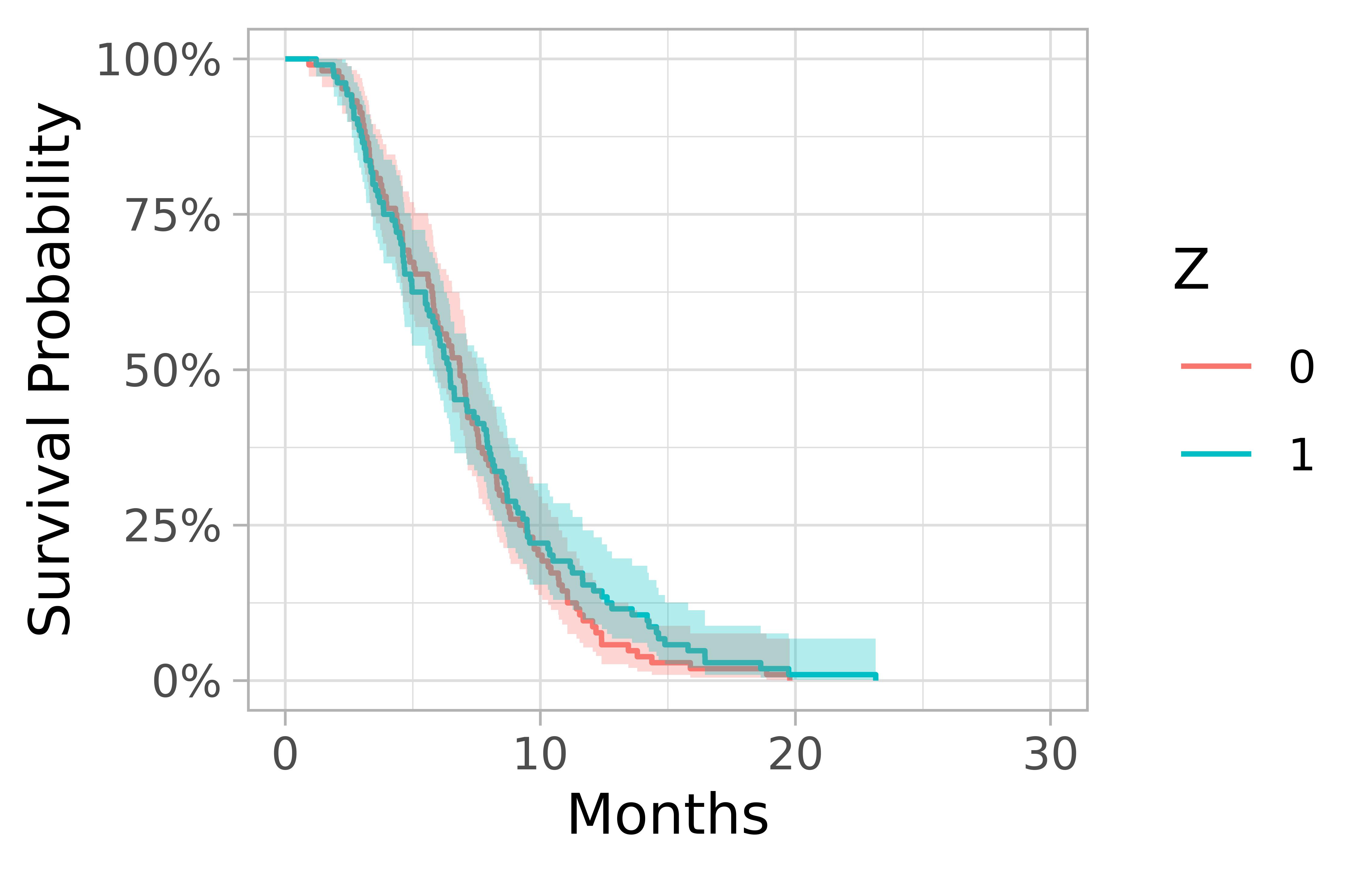}
  \caption{I$^{\text{ND}} = 0$}
  \label{kapmei2_D}
\end{subfigure}
\caption{\textit{Scenario II}. Kaplan-Meier curves for treated (blue) and controls (red), estimated using the potential outcomes' complete data (one sample).}
\label{kapmei2_ID}
\end{figure}

We simulate the data such that the principal causal effects for D patients are zero. 
However, the treatment has a positive effect on ND patients; the overall ITT effect is positive.
To mimic such a situation, as in Scenario I, we simulate the principal stratum membership and the time-to-discontinuation as in Equations \eqref{psmemberI} and \eqref{timetoDI}, respectively, and the potential outcomes for ND patients as in Equations \eqref{sc1_nd1} and \eqref{sc1_nd0}. 
Then, we generate the potential outcomes under investigational treatment and under control for D patients under the assumption that they are independent and identically distributed conditionally on the potential time to discontinuation and the covariates. 
Specifically, we assume that they both follow the following truncated Weibull distribution:
\begin{equation} \label{y10_sc2}
     \{Y_i(1)\mid \text{I}^{\text{ND}}_i = 0, D(1), X_i\} , \{Y_i(0)\mid \text{I}^{\text{ND}}_i = 0, D(1), X_i\} \overset{iid}{\sim} \text{tWeibull}_{D_i(1)}(\alpha, e^{-(\beta + X_i'{\eta} + \delta \log(D_i(1)))/\alpha})
\end{equation}
Here, the two potential outcomes have the same conditional distribution for each D patient.
The Kaplan-Meier curves based on potential outcomes' complete data (Figures \ref{kapmei2_ND} and \ref{kapmei2_D}) clearly show that there is no effect for the D patients; however, it is impossible to grasp that difference without resorting to the principal stratum strategy. 
Indeed, the observed survival curves in Figure \ref{kapmei2} show a positive ITT effect.
Further details on the data-generating processes are shown in the Supplementary Material. 

\begin{figure}[t]
    \centering
    \includegraphics[width=0.5\linewidth]{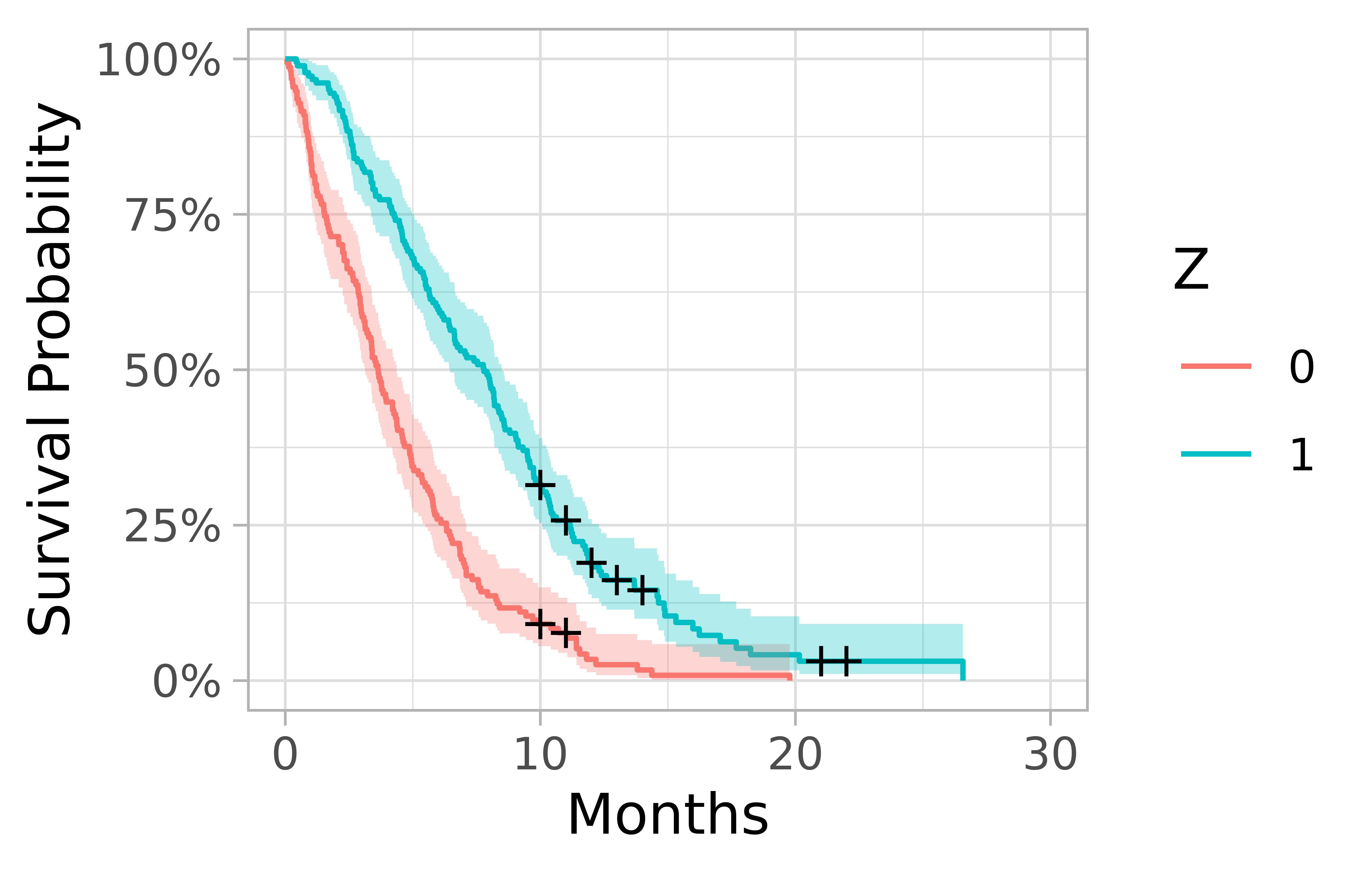}
    \caption{\textit{Scenario II}. 
    Kaplan-Meier curves for treated (blue) and controls (red), estimated using the observed data (one sample).}
    \label{kapmei2}
\end{figure}

\subsection{Modeling details and prior specification} \label{model_details}
We correctly specify the model for the principal stratum membership and the discontinuation time. 
This amounts to assuming that $\text{I}^{\text{ND}}_i$ follows a Bernoulli distribution as in Equation \eqref{psmemberI}, and the potential discontinuation under treatment, $D_i(1)\mid I_i^{\text{ND}} = 0$, is a Weibull as in Equation \eqref{timetoDI}; thus, $\theta^{\bar{\mathbb{D}}} = (\gamma_0,{\gamma} \in \mathbb{R}^{K+1})$, and $\theta^D = (\alpha_D, \beta_D, {\eta}_D), \alpha_D \in \mathbb{R}^+$, $\beta_D \in \mathbb{R}$, ${\eta}_D \in \mathbb{R}^K$.

Then, independently of how we simulated data, to estimate the causal effects under any scenario, we model the potential outcomes as follows:
\begin{equation}
    \psi_{\bar{1}}(\cdot; \bar{\theta}^1, X_i) = \text{Weibull}(\bar{\alpha}_1, e^{-(\bar{\beta}_1 + X_i'\bar{{\eta}}_1)/\bar{\alpha}_1} ), \quad \bar{\theta}^1 = (\bar{\alpha}_1, \bar{\beta}_1, \bar{{\eta}}_1)
\end{equation}
\begin{equation} 
    \psi_{\bar{0}}(\cdot; \bar{\theta}^0, X_i) = \text{Weibull}(\bar{\alpha}_0, e^{-(\bar{\beta}_0 + X_i'\bar{{\eta}}_0)/\bar{\alpha}_0} ), \quad \bar{\theta}^0 = (\bar{\alpha}_0, \bar{\beta}_0, \bar{{\eta}}_0)
\end{equation}
\begin{equation} \label{eq_tW}
    \psi_{1}(\cdot; \theta^1, X_i, D_i(1)) = \text{tWeibull}_{D_i(1)}(\alpha_1, e^{-(\beta_1 + X_i'{\eta}_1 + \delta\log(D_i(1)))/\alpha_1}), \; \theta^1 = (\alpha_1, \beta_1, {\eta}_1, \delta)
\end{equation}
\begin{equation}    \label{eq_miss}
    \psi_{0}(\cdot; \theta^0, X_i, D_i(1)) = \text{Weibull}(\alpha_0, e^{-({\beta}_0 + X_i'{\eta}_0+ \delta \log(D_i(1)))/\alpha_0}), \quad \theta^0 = (\alpha_0, \beta_0, {\eta}_0, \delta)
\end{equation}

The above model is a correct specification of the model under Scenario I. 
Under Scenario II, the model is misspecified due to Equation \eqref{eq_miss}: a truncated Weibull, rather than the Weibull, is the model we used to simulate data, and the parameters of the model are the same as those in Equation \eqref{eq_tW}.
The parameter $\delta$ in Equations \eqref{eq_tW} and \eqref{eq_miss} capture the dependence between the potential outcomes and the potential discontinuation time under treatment.
Letting $Y_i(1)$ and $Y_i(0)$ depending on the common parameter $\delta$ \citep[similarly as in][]{mattei2024assessing} makes the dependence between $Y_i(0)$ and $D_i(1)$ parametrically identifiable; the information on such a relation does not solely come from the prior.

Concerning the prior specification, we assume wide multivariate Normal priors for all parameters playing the role of covariates' coefficients and intercepts; we assume weakly informative Gamma priors for Weibulls' shape parameters.
Further details and the hyperparameters specification can be found in the Supplementary Material.

\subsection{Simulation study}\label{cov1}
We simulate 150 samples under each scenario described in Section \ref{synthetic} to evaluate the performance of our method in repeated sampling in terms of coverage and bias.
We estimate the model using the specification in Section \ref{model_details}, irrespective of the generating mechanism, setting the number of iterations equal to 50000, with a burnin of 30000 and a thinning of 5. 
To observe if a precision loss occurs in the absence of predictive covariates, we also estimate the model without using them.

Table \ref{coverage} shows how often the 95\% Highest Posterior Density intervals (HPD) cover the simulated values of the effects, i.e., the values computed using the complete simulated data, over 150 samples.
Table \ref{bias} shows the results for our estimates' bias, computed as the mean over 150 samples of the difference between the effects' posterior mean and their simulated values.
Each table shows the results for both scenarios when the model is estimated with or without including covariates.



\begin{table}[ht]
\centering
\caption{95\% HPD coverage with respect to the simulated values expressed as a relative frequency over 150 samples. }
\label{coverage}
\begin{tabular}{lcccccccc}
  \hline
 Scenario & $ITT$ & $ACE_{\text{ND}}$& $ACE_\text{D}$ & $ACE_\text{D}(1)$ & $ACE_\text{D}(2)$ & $ACE_\text{D}(3)$ & $ACE_\text{D}(4)$ \\  
  \hline
I & 0.96 & 0.95 & 0.95 & 0.94 & 0.94 & 0.95 & 0.95 \\ 
  I w/o covariates & 0.95 & 0.96 & 0.99 & 0.99 & 0.99 & 0.99 & 0.99 \\ 
  II & 0.93 & 0.79 & 0.69 & 0.97 & 0.92 & 0.73 & 0.39 \\ 
  II w/o covariates & 0.94 & 0.79 & 0.84 & 0.98 & 0.97 & 0.87 & 0.56 \\ 
   \hline
\end{tabular}
\end{table}

\begin{table}[ht]
\centering
\caption{Bias computed as the difference between the posterior mean and the simulated values, mean over 150 samples. Months.}
\label{bias} 
{\begin{tabular}{lcccccccc}
  \hline
 Scenario & $ITT$ & $ACE_\text{ND}$ & $ACE_\text{D}$ & $ACE_\text{D}(1)$ & $ACE_\text{D}(2)$ & $ACE_\text{D}(3)$ & $ACE(4)$  \\ 
  \hline
I & 0.06 & 0.26 & -0.38 & -0.31 & -0.26 & -0.24 & -0.24 \\ 
  I w/o covariates & 0.04 & 0.30 & -0.59 & -0.97 & -0.75 & -0.62 & -0.54 \\ 
  II & -0.16 & -0.88 & 1.68 & 0.77 & 1.17 & 1.80 & 2.54 \\ 
  II w/o covariates & -0.20 & -1.21 & 2.10 & 0.97 & 1.51 & 2.23 & 3.02 \\
  \hline
\end{tabular}}
\end{table}

The coverage in terms of $ITT$ and its bias are good. 
In terms of principal effects, under a correct model specification, the coverage is also overall good (Scenario I: Table \ref{coverage}, row 1), the bias is small for the ND effect (0.19 months is equivalent to 5.7 days; Table \ref{bias}, row 1) and still moderate for the D effects (in absolute value, 0.45 months, which is less than two weeks; Table \ref{bias}, row 1).
As expected, both the coverage and the bias worsen if the model is misspecified (Tables \ref{coverage} and \ref{bias}, rows 3). 
In this case, the longer the time to discontinuation, the worse the model's performance. 
The bias, which is equal to 25.8 and 44.4 days for the $ACE_{\text{ND}}$ and $ACE_\text{D}$ respectively, becomes larger as the time to discontinuation increases
due to how we specify the potential outcomes. 
In fact, both $Y(0)$ and $Y(1)$ were generated using a Weibull truncated on the time to discontinuation, whereas for estimation, we assume a Weibull without truncation for $Y(0)$. 
As the discontinuation time increases, the lower bound of the truncated Weibull becomes larger, and the expected value of $Y(0)$ becomes increasingly distant from that of its version without truncation. 
This leads to the natural bias that emerges from Table \ref{bias}. 
To overcome this problem, one could generally opt for more flexible models, e.g., mixture models. 
In our case, we find it sufficient to stick to a simpler model - with lower computational cost - because the time to discontinuation that we observe in the data is short: 50\% of patients discontinue within 1.4 months, 75\% within 4.1 (see Table \ref{novartis_synth}).

In any case, the results derived conditioning on the covariates under both Scenarios are superior to those obtained estimating the model without covariates. 
Table \ref{coverage} shows how the uncertainty around the estimates is much larger when covariates are not included; the 95\% HPD coverage sometimes exceeds 99\%, highlighting a bad approximation of the tails of the posterior distribution of the effect. 
Hence, as expected, the bias in the absence of covariates increases (Table \ref{bias}, rows 2 and 4).
Intuitively, proper utilization of auxiliary variables provides extra dimensions to predict the missing principal strata membership better; recent results on mixture models show that a multivariate analysis improves the efficiency of estimators \citep{mercatanti2015improving}. 
In this case, including covariates may mitigate the impact of the model misspecification.

These results show how crucial it is to include available covariates that are also good predictors of the potential outcomes and latent principal stratum membership in the context of treatment discontinuation in RCTs.

In the next subsection, we show the model's performance on one data set per scenario.

\subsection{Estimation of causal effects}\label{sec:scenarios}
\subsubsection{Scenario I}

Table \ref{tab_scenario1} shows the results for the principal average causal effects and the ITT effect computed as the weighted average of the principal $ACE$'s. 
The posterior means of the effects are all positive.

\begin{table}[ht]
\centering
\caption{\textit{Scenario I}. Posterior mean and 95\% HDP interval of the percentage of ND patients and the population causal estimands considered.} 
\label{tab_scenario1}
{\begin{tabular}{lccc}
\hline
 & Posterior Mean &\multicolumn{2}{c}{95\% HPD} \\ \hline
 $\pi_\text{ND}$ & 0.66 & \multicolumn{2}{l}{[0.61 ; 0.70]}
 \\ \hline
 $\pi_{\text{ND}}ACE_{\text{ND}}+(1-\pi_{\text{ND}})ACE_{\text{D}}$ & 4.66 & \multicolumn{2}{c}{[3.14 ; 5.19]} \\ 
$ACE_{\text{ND}}$ & 4.33 & \multicolumn{2}{c}{[2.88 ; 5.97]} \\ 
$ACE_{\text{D}}$ & 3.92 & \multicolumn{2}{c}{[2.17 ; 5.53]} \\ 
\hline
\end{tabular}}
\end{table}

\begin{figure}[t]
    \centering
        \includegraphics[width=.6\linewidth]{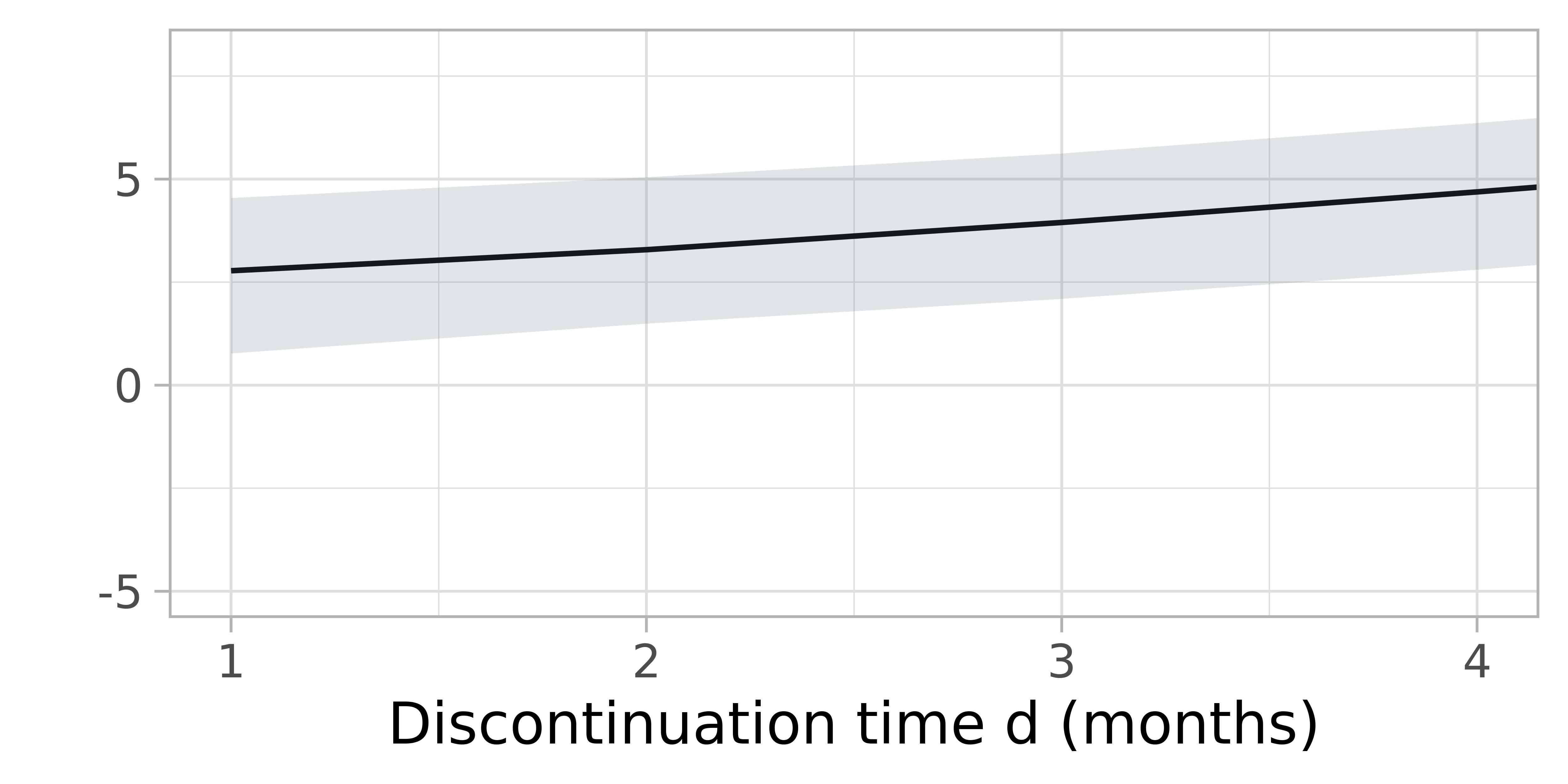}
    \caption{\textit{Scenario I}. $ACE_{\text{D}}(d)$ as a function of the potential discontinuation time $D(1) = d$. 
    Posterior mean (dotted line) and 95\% HPD interval (grey shade).}
    \label{ACEd}
\end{figure}

Figure \ref{ACEd} shows the population average causal effect for D patients as a function of the discontinuation time.
Patients who receive the treatment for longer may benefit more from the new investigational drug even though they will experience an AE at a certain time.

\begin{figure}[t]
    \centering
        \includegraphics[width=.6\linewidth]{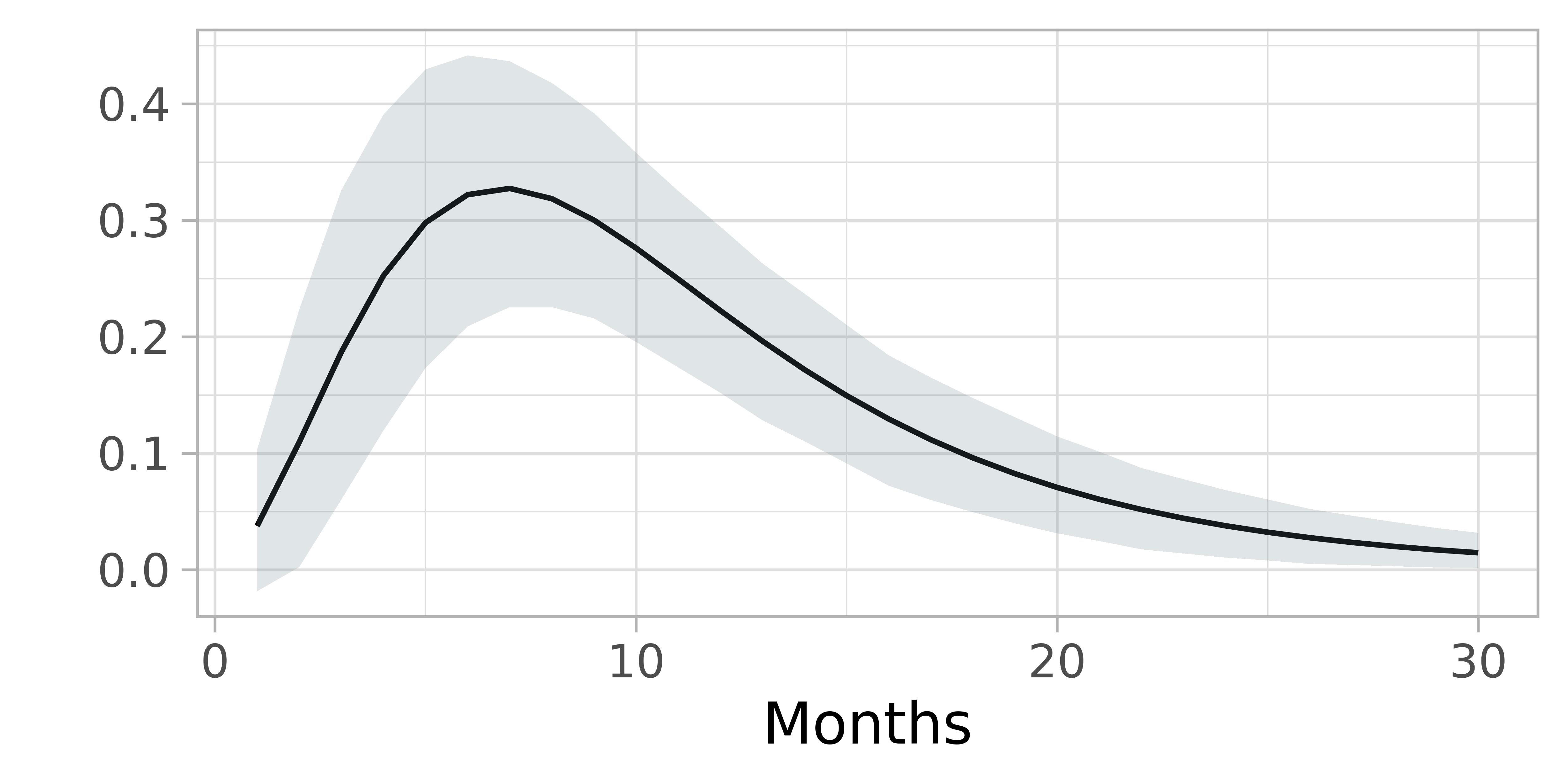}
    \caption{\textit{Scenario I}. Principal survival difference at time $y$ of ND patients, $DCE_\text{ND}$.
    Posterior mean (solid line) and 95\% HPD interval (grey shade).}
    \label{DCEnd}
\end{figure}

\begin{figure}[t]
    \centering
        \includegraphics[width=.7\linewidth]{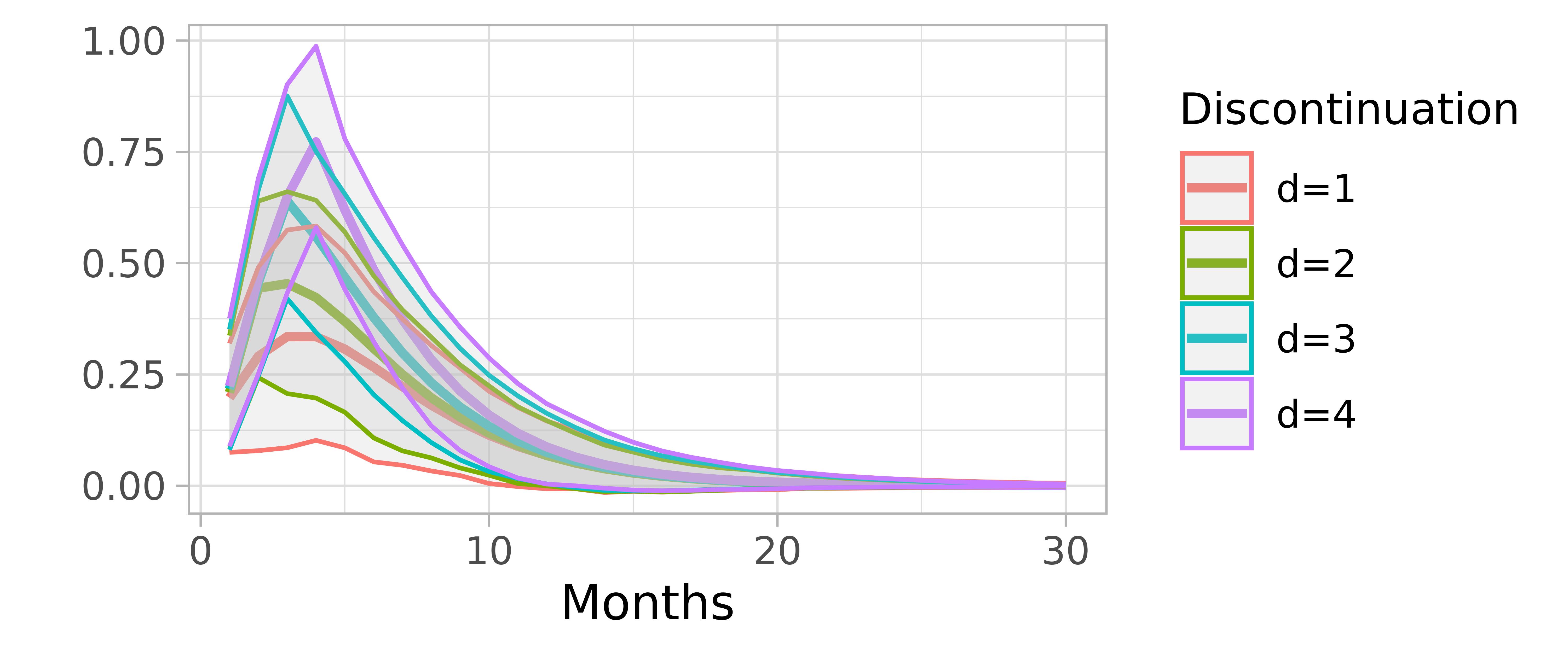}
    \caption{\textit{Scenario I}. Principal survival difference at time $y$ of D patients, $DCE_\text{D}(y\mid d)$, for different potential discontinuation time $D(1) = d$ (in different colours).}
    \label{DCEd}
\end{figure}

As stated in Section \ref{sec:postcomp}, this Bayesian PS approach allows us to draw values from the posterior distribution of any valid causal estimand of interest beyond the $ACE$'s, such as the distributional survival differences causal effects of the treatment defined in Equations \eqref{dce_nd} and \eqref{dce_d}. 
\color{black}
Figures \ref{DCEnd} and \ref{DCEd} show the principal survival differences. 
As we expect, both $DCE_{\text{ND}}(y)$ and the $DCE_\text{D}(y\mid d)$ for each $d$ are first concave, and then, after the peak, they flex and become convex. 
Such behavior indicates that the survival curve under control decreases much faster than the one under treatment.

\subsubsection{Scenario II}
Table \ref{tab_scenario2} summarises the results obtained under Scenario II in terms of principal causal effects and ITT effect computed as the weighted average of the principal $ACE$'s.
The overall ITT effect is positive; the principal stratification approach allows us to highlight that the stratum of ND patients leads to such a result.
In fact, the $ACE_{\text{ND}}$ is positive, whereas we cannot reject the hypothesis of no causal effect for D patients since the 95\% HPD interval of $ACE_{\text{D}}$ covers the zero. 

\begin{table}[ht]
\centering
\caption{\textit{Scenario II}. Posterior mean and 95\% HPD interval of the percentage of ND patients and the population causal estimands considered.} 
\label{tab_scenario2}
{\begin{tabular}{lccc}
\hline
  & Posterior Mean &\multicolumn{2}{c}{95\% HPD} \\ \hline
 $\pi_\text{ND}$ & 0.67 & \multicolumn{2}{l}{[0.63 ; 0.72]}
 \\ \hline
 $\pi_{\text{ND}}ACE_{\text{ND}}+(1-\pi_{\text{ND}})ACE_{\text{D}}$ & 3.84 & \multicolumn{2}{c}{[2.85 ; 4.89]} \\ 
$ACE_{\text{ND}}$ & 5.85 & \multicolumn{2}{c}{[3.46 ; 6.72]} \\ 
$ACE_{\text{D}}$ & 1.14 & \multicolumn{2}{c}{[-0.89 ; 3.19]} \\ 
\hline
\end{tabular}}
\end{table}

\begin{figure}[t]
    \centering
    \includegraphics[width=0.6\linewidth]{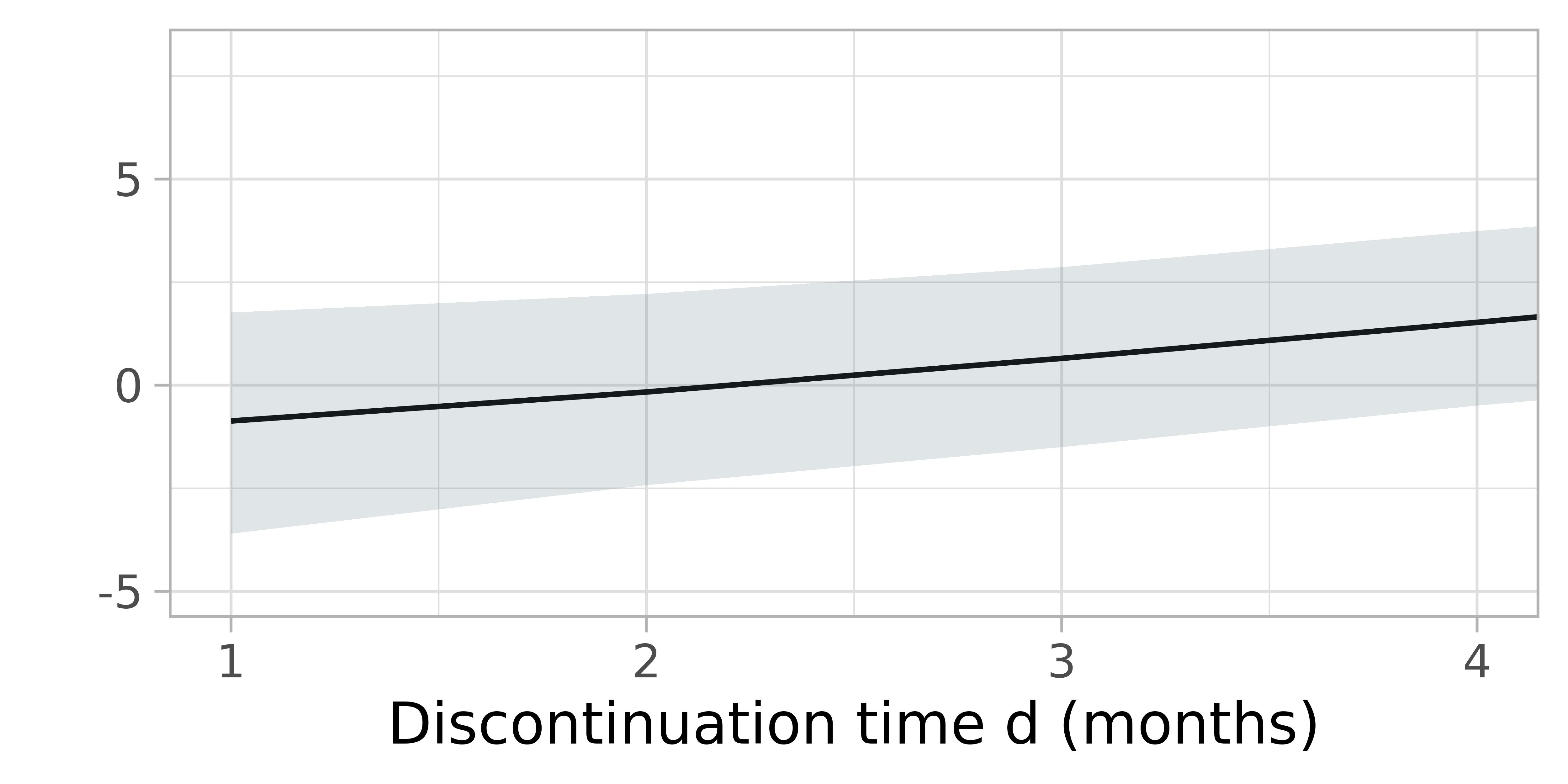}
    \caption{\textit{Scenario II}. $ACE_\text{D}(d)$, as a function of the potential discontinuation time $D(1) = d$. 
    Posterior mean (solid line) and 95\% HPD interval (grey shade).}
    \label{ACEd_2}
\end{figure}

The misspecification of the model leads to more bias as the discontinuation time gets longer (see Figure \ref{ACEd_2}).
However, we remind that our simulated data mimic the data in Table \ref{novartis_synth}, where about 75\% of patients discontinue within the first 4 months. 
Yet, the $ACE_{\text{D}}(d)$ HPD interval at the 95\% credibility level shown in Figure \ref{ACEd_2} is widely covering the 0. 
Looking at the HPD intervals for the $ACE_\text{D}(d)$, we can notice how they are much wider under Scenario II than under Scenario I; this may be due to a propagation of the misspecification error. 

\begin{figure}[t]
    \centering
    \includegraphics[width=0.6\linewidth]{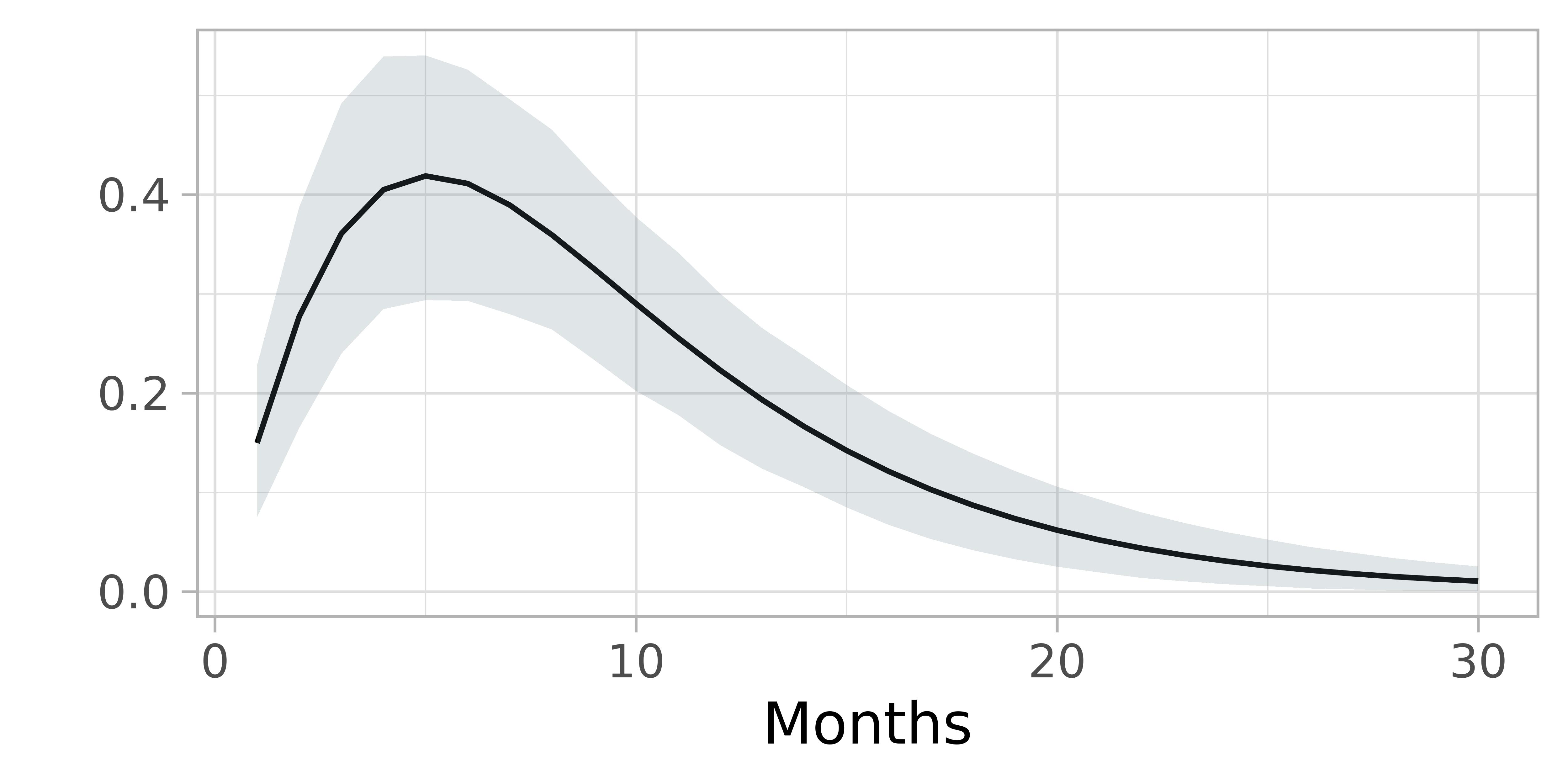}
    \caption{\textit{Scenario II}. Principal survival difference of ND patients, $DCE_\text{ND}$.
    Posterior mean (dotted line) and 95\% HPD interval (grey shade).}
    \label{DCEnd_2}
\end{figure}

\begin{figure}[t]
    \centering
    \includegraphics[width=.7\linewidth]{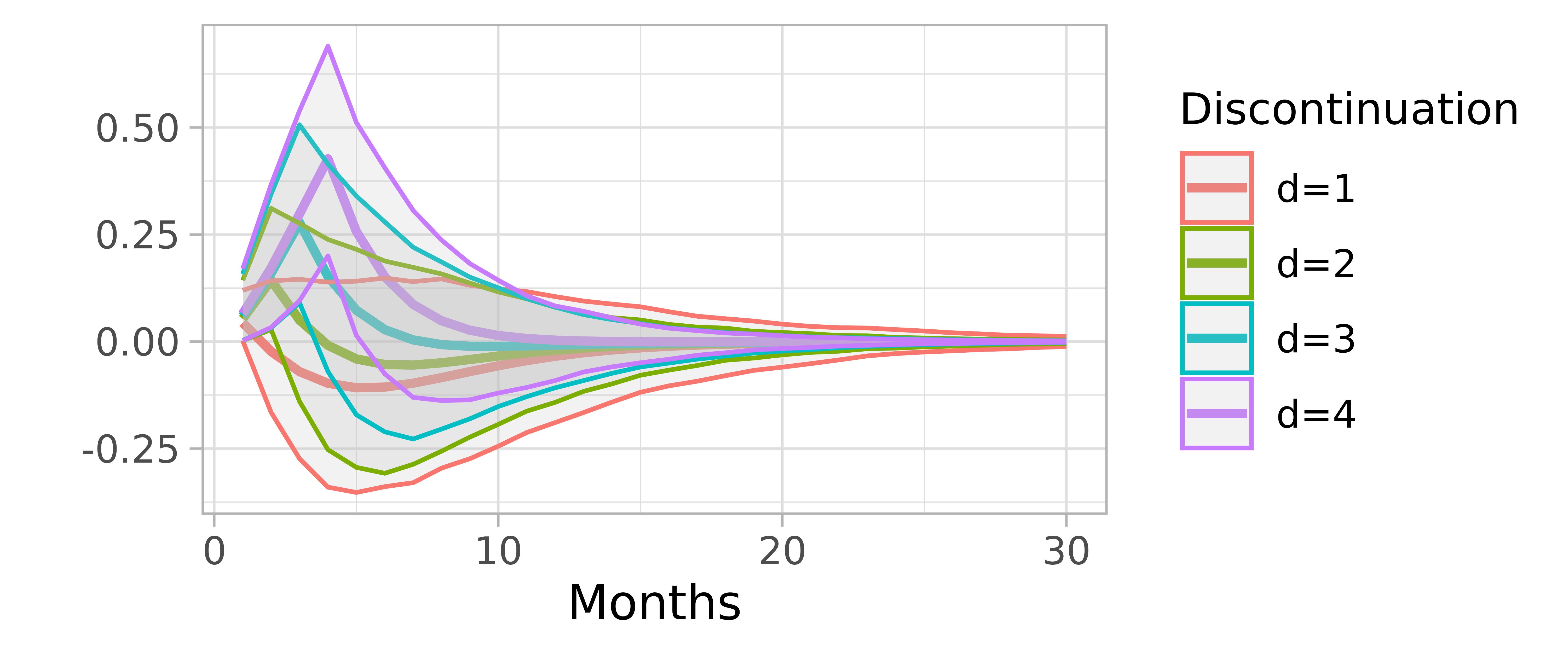}
    \caption{\textit{Scenario II}. Principal survival difference at time $y$ of D patients, $DCE_\text{D}(y\mid d)$, for different potential discontinuation time $D(1) = d$ (in different colours).}
    \label{DCEd_2}
\end{figure}

Figures \ref{DCEnd_2} and \ref{DCEd_2} show the principal survival differences for ND and D patients, respectively. 
As for the principal $ACE$'s, there is evidence of a positive treatment effect on the progression-free survival for ND patients.
Indeed, as in Scenario I, $DCE_\text{ND}(y)$ shows a peak and then slowly goes to zero. 
However, here, the HPD intervals are much wider, indicating more uncertainty deriving from the misspecification of the model. 
$DCE_{\text{D}}(y\mid d)$ sharply go to zero, suggesting there is no significant treatment effect for the D patients.

\section{The role of covariates: Characterization of the principal strata} \label{cov2}
We previously stressed the role of covariates as predictors that allow more robust estimates of the causal effects in the simulation study in Section \ref{cov1}.
The results are in line with the literature.
In a principal stratification analysis, even under randomization, the use of covariates improves inference by helping the prediction of missing potential outcomes, and thus the identification of the principal causal effects \citep[see][]{gilbert2008evaluating,grilli2008nonparametric, ding2011identifiability,long2013sharpening,mercatanti2015improving,mealli2013using,jiang2021identification}.

Here, we highlight the importance of the covariates from a different point of view, emphasizing how our approach may inform about the risk of AEs and how it makes the characterization of the patients' behavior natural in terms of the patient's baseline features.
Indeed, we may be interested in profiling the patients in the different principal strata in terms of the covariates so we can better predict whether and when a patient will discontinue the treatment.
For a likelihood-based approach to this issue, see \cite{frumento2012evaluating}.

We investigate the distribution of the covariates within different principal strata, e.g., ND patients, \textit{early} D patients, and \textit{late} D patients. 
We arbitrarly define ``early D patients'' as those patients discontinuing before the observed median time of discontinuation, whereas with ``late D patients'' we indicate those discontinuing after the observed median.

\begin{figure}[t]
    \centering
    \includegraphics[width=0.6\linewidth]{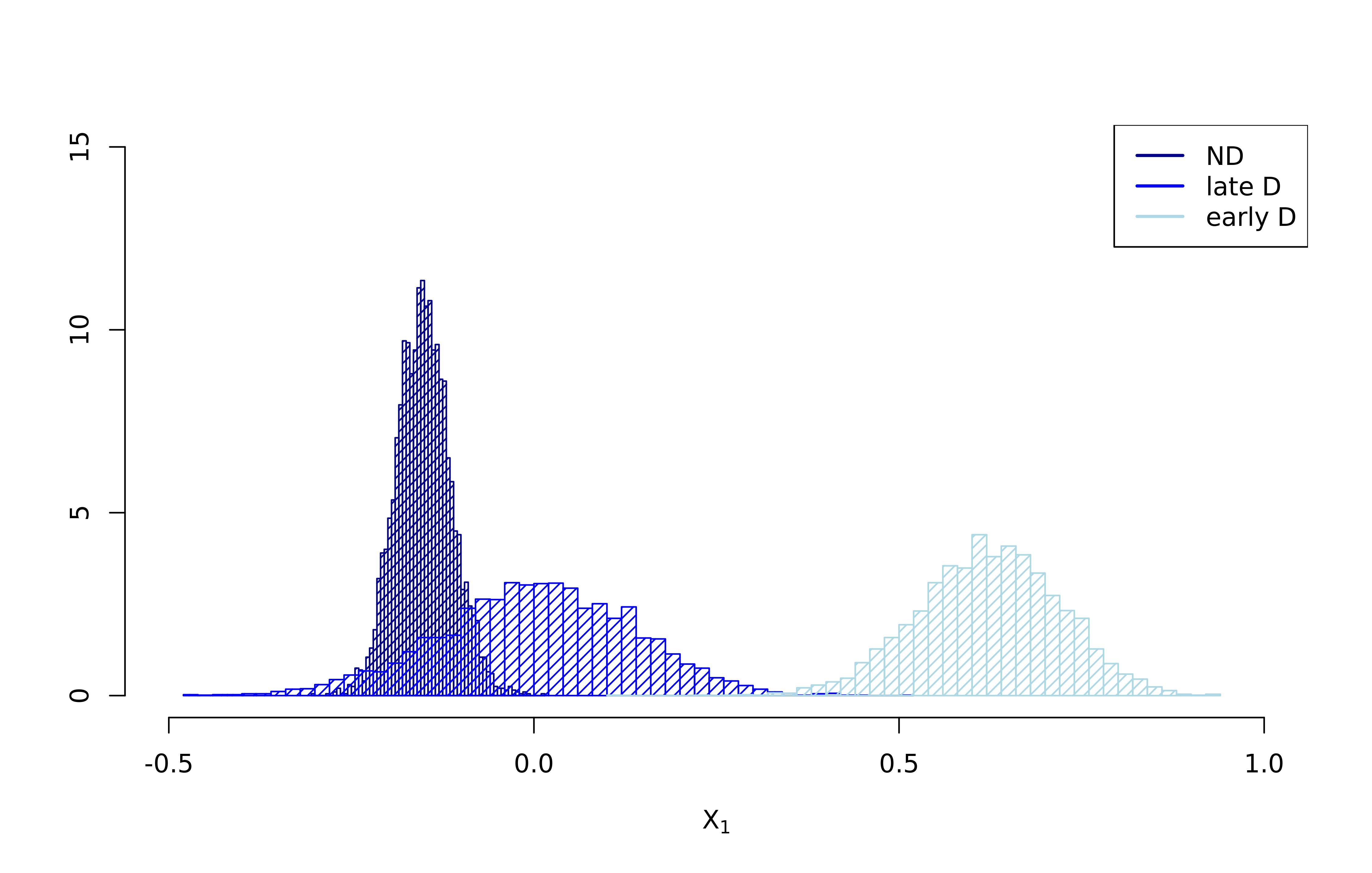}
    \caption{Distribution of $X_1$ posterior mean by latent stratum - ``ND'' refers to ND patients; ``late D'' and ``early D'' refer to D patients whose $D_i(1) \geq \texttt{median} (D_i(1))$ and $D_i(1) < \texttt{median} (D_i(1))$, respectively.}
    \label{x1}
\end{figure}

\begin{figure}[t]
    \centering
    \includegraphics[width=0.6\linewidth]{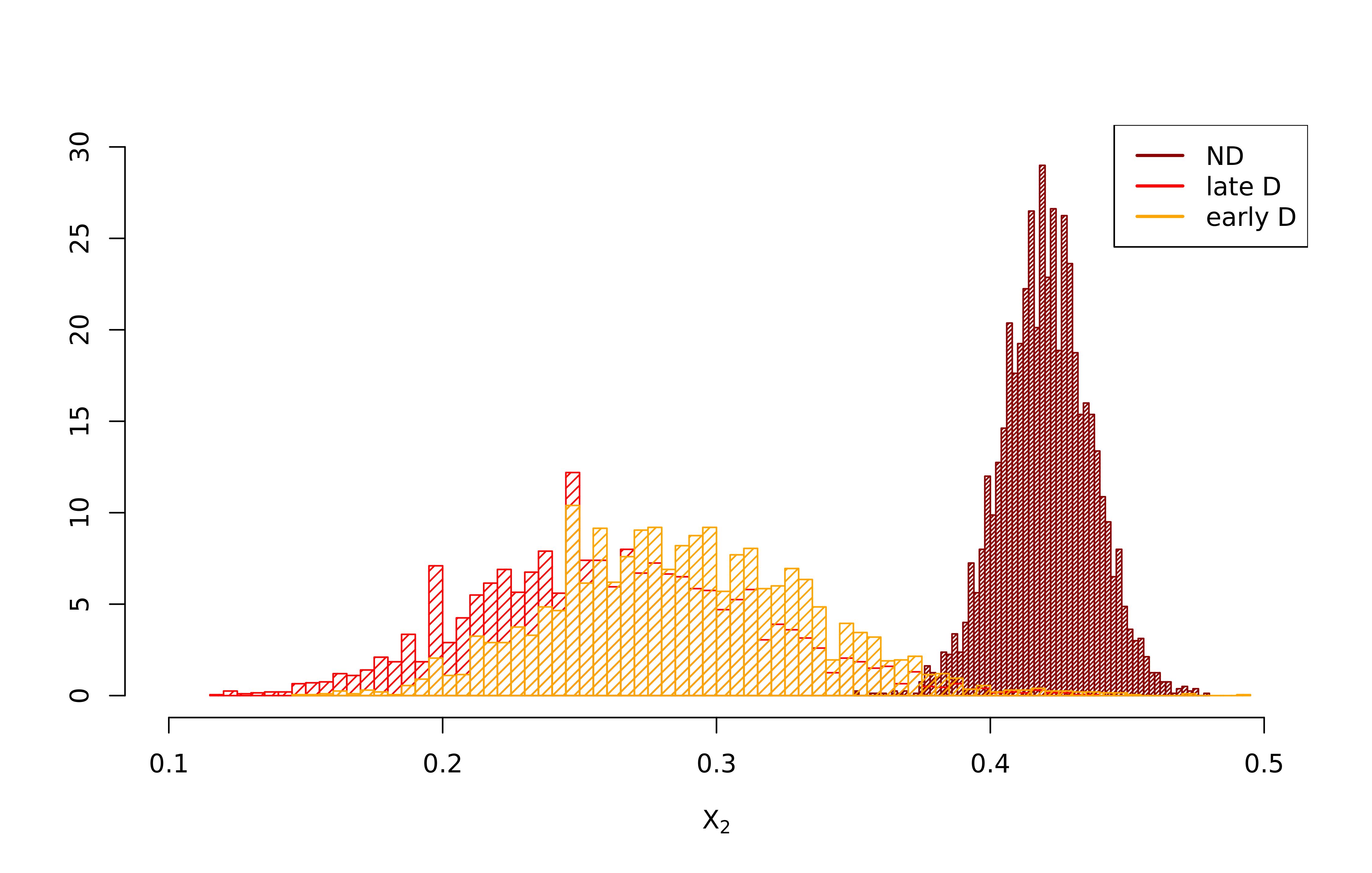}
    \caption{Distribution of $X_2$ posterior mean by latent stratum - ``ND'' refers to ND patients; ``late D'' and ``early D'' refer to D patients whose $D_i(1) \geq \texttt{median} (D_i(1))$ and $D_i(1) < \texttt{median} (D_i(1))$, respectively.}
    \label{x2}
\end{figure}

\begin{figure}[t]
    \centering
    \includegraphics[width=0.6\linewidth]{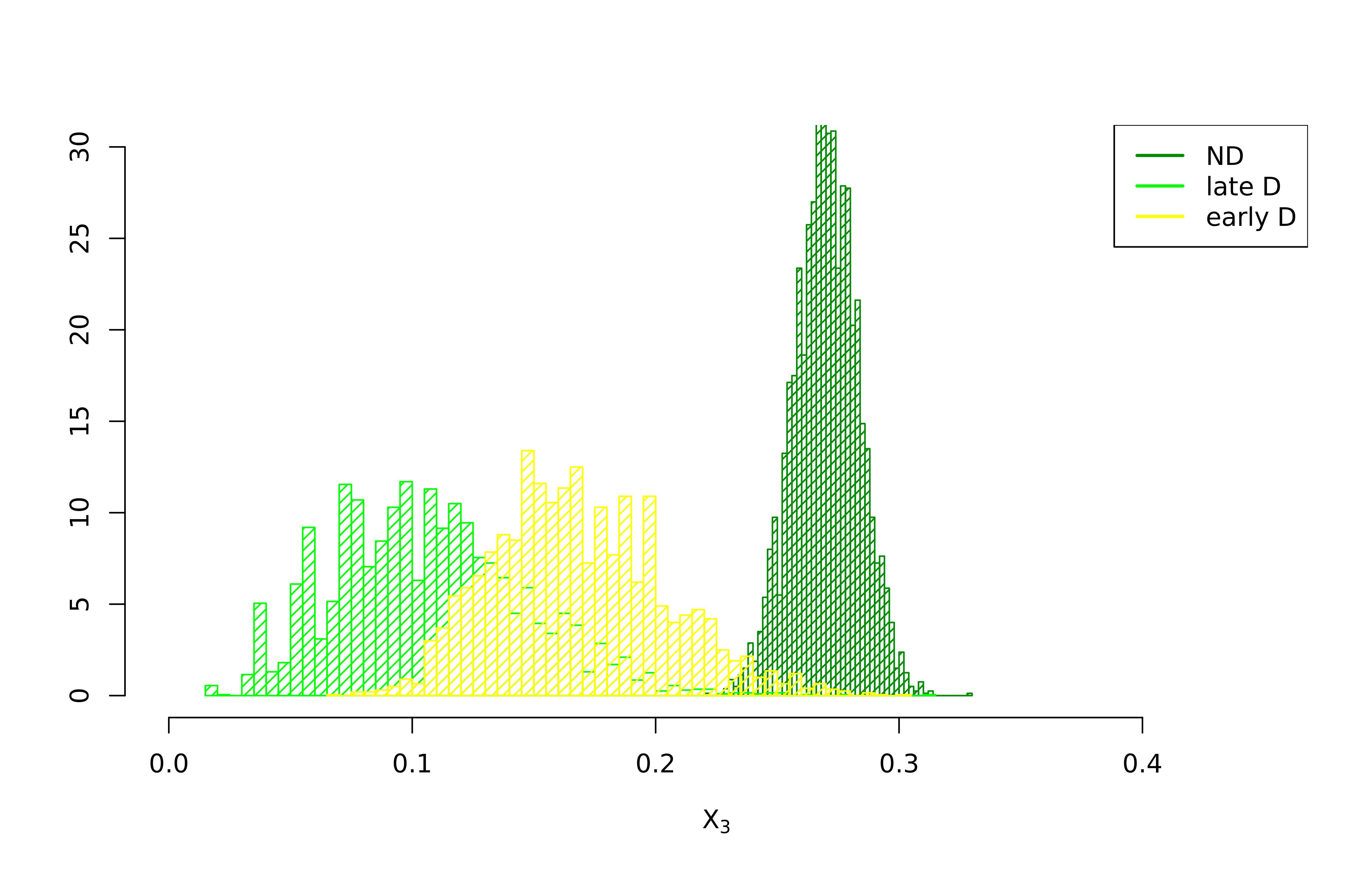}
    \caption{Distribution of $X_3$ posterior mean by latent stratum - ``ND'' refers to ND patients; ``late D'' and ``early D'' refer to D patients whose $D_i(1) \geq \texttt{median} (D_i(1))$ and $D_i(1) < \texttt{median} (D_i(1))$, respectively.}
    \label{x5}
\end{figure}

The availability of posterior samples of memberships and discontinuation time make it straightforward to characterize the principal strata. 
Figures \ref{x1}-\ref{x5} show the distributions of the posterior means of the covariates used in our analyses with respect to the patients' membership.

ND patients are characterized by lower values of $X_1$ (Figure \ref{x1}).
However, given that one would discontinue the treatment, the lower the value of $X_1$, the later the discontinuation.

The higher the probability that $X_2$ or $X_3$ equals 1, the more likely the patient would be an ND one. 
However, if one discontinues, those with a higher probability of $X_3 = 1$ discontinue sooner; the association of $X_2$ to the discontinuation time is not strong.

The results are reasonable and consistent with the assumptions used in simulating the data.

\section{Discussion}
This paper addresses a relevant but challenging clinical question. 
A novel treatment (administered in combination with the SOC) is more likely to trigger AEs, leading to discontinuation of the novel component while continuing the SOC alone.
Suppose a trial shows a clinically meaningful overall treatment effect based on the ITT analysis. 
Naturally, one may be interested in the treatment effect in those who adhere to the treatment but also in those who discontinue the treatment.
This would help better understand the drug mechanism (e.g., whether taking only some doses initially may prove a longer-lasting benefit even after early discontinuation) and provide a more complete picture for clinical decision-making. 
For instance, whether early treatment discontinuation leads to reduced or null treatment benefit, whether treatment discontinuation depends on specific baseline covariates, and so on, all constitute clinically relevant information that can be used in the risk/ benefit assessment of a new treatment for an individual patient.
We adopt a principal stratification approach, classifying patients according to their potential discontinuation time under treatment.
Our approach enables us to decompose the ITT effect into the actual treatment effect for ND patients and the effects of initiating the treatment for D ones.
Hence, it allows us to evaluate whether, e.g., a positive ITT effect results from either positive effects for each type of patient or only for some specific subgroups.
We emphasize that, although other summaries of the causal effects are possible, we choose to express the results in terms of average causal effects and survival differences also because of the limitations that other estimands, such as the hazard ratio \citep{stensrud2019limitations}, may have.

For inference, we use a flexible parametric Bayesian model. 
In the motivating application, we used specific parametric models; however, the approach is more general and can also be implemented with alternative model specifications.
All model components allow for the inclusion of baseline covariates, which makes the parametric assumptions more plausible and improves the prediction of the missing potential outcomes. 
It also potentially allows us to characterize the D patients by discontinuation time and the ND patients in terms of baseline covariates.
This provides very useful and readily interpretable information, having substantial implications for further drug development: if AEs are associated with specific baseline characteristics and if patients who discontinue the treatment have a reduced treatment effect, one could investigate an optimized dosing regime or improved AEs management guideline for these patients in further research. 

In this work, we assume that covariates that are good predictors of outcome variables (time to event, discontinuation status, and time to discontinuation) are available. 
In practice, covariates or prognostic factors impacting time to progression or death are usually well-discussed, recognized, and collected for specific disease areas. 
In contrast, baseline covariates that may impact the discontinuation status may not always be clear and readily available, especially for rare diseases and novel treatments (as well as its chosen dosing regimen).
Hence, at the design stage, it is essential to assess potential covariates (based on earlier trials or clinical or mechanistic understanding) and make them part of the data collection.
Our simulation study shows that the model performance strongly improves by including appropriate covariates in the analysis.

\section{Supplementary Material}\label{sec6}
Supplementary Material includes:
\begin{enumerate}
    \item the details of the synthetic data generating process;
    \item the specification of the model used in the estimation (specification of density functions and survival functions);
    \item details about priors' specification;
    \item the derivation of the posterior using the observed data likelihood and the complete data posterior distribution;
    \item the complete MCMC algorithm with data augmentation;
    \item two Tables showing the summary statistics under the two scenarios described in this article.
\end{enumerate}

\section*{Acknowledgments}
Veronica Ballerini is supported by the European Union - Next GenerationEU, UNIFI Young Independent Researchers Call - BayesMeCOS Grant no. B008-P00634. 
\\
Alessandra Mattei and Fabrizia Mealli are supported by the Italian Ministry of University and Research (MUR), Department of Excellence project 2023-2027 ReDS `Rethinking Data Science' - Department of Statistics, Computer Science, Applications - University of Florence. \\
{\it Conflict of Interest}: None declared.

\bibliographystyle{abbrvnat}
\bibliography{main}

\clearpage
\appendix

\section*{\huge Supplementary material}

This file contains:
\begin{enumerate}
    \item the details of the synthetic data generating process, Section \ref{app0};
    \item details about priors' specification, Section \ref{app2};
    \item the derivation of the posterior using the observed data likelihood and the complete data posterior distribution, Section \ref{app3};
    \item the complete MCMC algorithm with data augmentation, Section \ref{app5};
    \item two Tables showing the summary statistics under the two scenarios described in this article, Section \ref{app6}.
\end{enumerate}

\section{Synthetic data generating process} \label{app0}
We use information in Tables 1 and 2, Section 2, of the main text to generate individual data under different scenarios (150 samples). 

The two scenarios described in Section 5.1 of the main text differ only in the assumptions made on the treatment effect; thus, only the potential outcomes-generating process will differ. 

We fix the sample size $n = 335$, where $n_T = 181$ is the number of treated units and $n_C = n - n_T = 154$ is the number of controls. 
$Z$ is a vector concatenating $n_T$ 1's and $n_C$ 0's.

\subsection{Covariates}
Recall that the baseline information consists of the following covariates:
\begin{itemize}
    \item $X_1$: continuous variable. The higher the value, the higher the risk of progression.
    \item $X_2$: binary indicator. $X_2$ is associated with a higher progression risk.
    \item $X_3$: binary indicator. $X_3$ is associated with a higher progression risk.
\end{itemize}

We randomly generate $n$ values for the three covariates as follows:
$$x_{i,1} \sim \text{N}(\mu = 63.27, \sigma = 10.5)\;, \quad i = 1, \dots, n\;;$$
$$x_{i,2} \sim \text{Ber}(\pi_2 = 0.44)\;, \quad i = 1, \dots, n\;;$$
$$x_{i,3} \sim \text{Ber}(\pi_3 = 0.24)\;,  \quad i = 1, \dots, n\;;$$
where the parameters' values are fixed based on the summary statistics in Table 1 of the main text. 
We standardise $x_1$ and use the standardised version both to generate data and in the estimation procedure. 

\subsection{Discontinuation}
We assume that whether a patient would discontinue the treatment or not if assigned to it depends on the intrinsic, pre-treatment, characteristics of that patient; thus, if one could observe patients until progression or death, the randomization would ensure that the proportion of patients who potentially discontinue under treatment is equal in the two treatment arms.
However, in our case study, the discontinuation time is censored by the administrative end of the study.

Hence, the proportion of patients who are observed to be ND patients, namely those who are observed to progress or die under treatment without discontinuing the treatment (90/181, Table 2, Section 2 of the main text), is a lower bound for the treated patients who discontinue the treatment; the upper bound is 151/181.
Therefore, we assume the proportion of ND patients in the whole sample to be between $.5$ and $.8$. 

Let $I^{\text{ND}}_i, i = 1, \dots, n$, be the indicator variable that takes value 1 if the patient $i$ is an ND patient, and $X_i$ be the 3-dimensional vector of covariates, $X_i = (X_{i,1},X_{i,2},X_{i,3})$.
We randomly draw $n$ values of $I^{\text{ND}}$ from a Bernoulli with probability parameter 
$$    p({X}_i) = \dfrac{\exp(\gamma_0 + {x}'_i {\gamma})}{1+\exp(\gamma_0 + {x}'_i {\gamma})}\;, \quad i = 1, \dots, n\; ,$$
where $\gamma_0 = .6$ so that $ p({X}_i) \approx .65$ when all covariates are set to $0$, and
\begin{itemize}
    \item $\gamma_1 = -0.5$; we assume the higher $X_1$, the lower the probability to be an ND patient;
    \item $\gamma_2 = 0.45, \gamma_3 = 0.55$; we assume that the more severe the condition, the higher the probability of being an ND patient.
\end{itemize}

Once the covariates and the basic strata (being an ND or a D patient) are generated, we generate the potential discontinuation time for all D patients, irrespective of their actual treatment assignment, drawing random values from Weibull distributions. 

Let the density function of a generic Weibull random variable $Y$ with a shape and scale parameterization, i.e., $Y \sim \text{Weibull}(a,b)$, be defined as:
\begin{equation}\label{weibull}
    f_Y(y\mid a, b) = a y^{a-1} b^{-a} \exp\{-y^ab^{-a}\} \; .
\end{equation}
Hence, potential discontinuation time $D_i(1)$ is generated as follows:
\begin{equation*}
    D_i(1) \mid {X}_i \sim \text{Weibull}(a = \alpha_D, b = e^{-(\beta_D+{X}'_i{\eta}_D)/\alpha_D})\; , \quad i:I_i^{\text{ND}} = 0\;.
\end{equation*}
We set the parameters' values as follows:
\begin{itemize}
    \item the shape parameter, $a = \alpha_D$, is set equal to 1.2, as a shape $>1$ is a typical choice for Weibull distributions describing time-to-event variables
    \item the scale parameter $b$ is expressed as a function of the covariates, i.e., $b = \exp\{-(\beta_D +
    {X}'_i {\eta}_D)/\alpha_D\}$. 
    \begin{itemize}
        \item[$\bullet$] The intercept $\beta_D$ is set such that the expected value of $D(1)$ in the absence of covariates and for $\alpha_D = 1.2$ matches the value in Table 1.
        \item[$\bullet$] Parameters $\eta_{D,1}, \eta_{D,2}, \eta_{D,3}$ are set equal to 0.45, 0.25, and 0.35, respectively; we assume that D patients with higher risk discontinue sooner.
    \end{itemize}
\end{itemize}

\subsection{Potential outcomes}
As described in Section 5.1 of the main text, under both scenarios, we generate the potential outcome under treatment $Y_i(1), i = 1, \dots, n$, given the discontinuation status $D_i(1)$ and the covariates ${X}_i$. 
For ND patients, we draw random values from a Weibull distribution parametrized as \eqref{weibull}, whereas for the D patients we generate values from a left truncated Weibull, truncated on the discontinuation time $D_i(1)$:
$$Y_i(1)|\text{I}^\text{ND}_i,D_i(1),{X}_i \begin{cases}
    \sim \text{Weibull}(\bar{\alpha}_1, e^{-(\bar{\beta}_1 +
     {X}'_i\bar{{\eta}}_1)/\bar{\alpha}_1}) & \text{if }  \text{I}^\text{ND}_i = 1\\
    \sim \text{tWeibull}_{D_i(1)}(\alpha_1, e^{-(\beta_1 
     + {X}'_i{\eta}_1+ \delta\log(D_i(1)))/\alpha_1}) & \text{if } \text{I}^\text{ND}_i = 0 \; .
    \end{cases}$$

We set the parameters' value for reasons similar to those described in the previous paragraph concerning the discontinuation time. 
In particular, in both scenarios:
\begin{itemize}
\item $\bar{\alpha}_1 = 1.7$; $\alpha_1 = 1.6$;
\item $\bar{\beta}_1$ and $\beta_1$ are set such that the expected value of $Y(1)$ for ND and D patients in the absence of covariates are equal to 11 and 7 months, respectively;
\item $\bar{{\eta}}_1 = {\eta}_1 = (0.25, 0.7, 0.45)$;
\item[$\bullet$] $\delta = 0.3$.
\end{itemize}

Yet, the potential outcome under control, $Y_i(0), i = 1, \dots, n$, is generated differently in the two scenarios.
In particular:
\begin{itemize}
\item[] Scenario I:
$$    Y_i(0)|\text{I}^\text{ND}_i, D_i(1),{X}_i \begin{cases}
    \sim \text{Weibull}(\bar{\alpha}_0, e^{-(\bar{\beta}_0 + {X}'_i\bar{{\eta}}_0)/\bar{\alpha}_0}) & \text{if }  \text{I}^\text{ND}_i = 1\\
    \sim \text{Weibull}(\alpha_0, e^{-(\beta_0 + {X}'_i{\eta}_0+ \delta\log(D_i(1)))/\alpha_0}) & \text{if } \text{I}^\text{ND}_i = 0 \; ,
    \end{cases}$$
    where
    \begin{itemize}
        \item[$\bullet$] $\bar{\alpha}_0 = 1.6$; $\alpha_0 = 1.5$;
        \item[$\bullet$] $\bar{\beta}_1$ and $\beta_1$ are set such that the expected value of $Y(1)$ for ND and D patients in the absence of covariates are equal to 5 and 4 months, respectively;
        \item[$\bullet$] $\bar{{\eta}}_1 = {\eta}_1 = (0.25, 0.7, 0.45)$; 
        \item[$\bullet$] $\delta = 0.3$.
    \end{itemize}
    \item[] Scenario II:
    $$    Y_i(0)|\text{I}^\text{ND}_i, D_i(1),{X}_i \begin{cases}
    \sim \text{Weibull}(\bar{\alpha}_0, e^{-(\bar{\beta}_0 + {X}'_i\bar{{\eta}}_0)/\bar{\alpha}_0}) & \text{if }  \text{I}^\text{ND}_i = 1\\
    \sim \text{tWeibull}_{D_i(1)}(\alpha_0, e^{-(\beta_0 + {X}'_i{\eta}_0+ \delta\log(D_i(1)))/\alpha_0}) & \text{if } \text{I}^\text{ND}_i = 0 \; ,
    \end{cases}$$
    \begin{itemize}
        \item[$\bullet$] $\bar{\alpha}_0 = 1.6$; $\alpha_0 = 1.6$;
        \item[$\bullet$] $\bar{\beta}_1$ and $\beta_1$ are set such that the expected value of $Y(1)$ for ND and D patients in the absence of covariates are equal to 4 and 7 months, respectively;
        \item[$\bullet$] $\bar{{\eta}}_0 = {\eta}_0 = (0.25, 0.7, 0.45)$; 
        \item[$\bullet$] $\delta = 0.3$.
    \end{itemize}
    \end{itemize}
\subsection{Censoring}
We randomise the follow-up time drawing the entry to the study from a mixture: the entry month is the last one, namely the 23$^{rd}$ month, with 50\% probability; otherwise, it is a random month between months 0 and 23. 

\section{Priors specification}\label{app2}
We assume multivariate Normal priors for all parameters playing the role of covariates' coefficients and intercepts; yet, we assume Gamma priors for Weibulls' shape parameters.

\begin{equation*} 
\theta^{\bar{\mathbb{D}}} := (\gamma_0, {\gamma}) \sim \text{N} ({0},\Sigma_{\bar{\mathbb{D}}})
\end{equation*}
\begin{equation*}
\theta^D := (\alpha_D, \beta_D, {\eta}_D), \quad \hbox{where}
\begin{cases}
\alpha_D \sim \text{Gamma}(a,b),  \\
(\beta_D, {\eta}_D) \sim \text{N}({0}, \Sigma_D)
\end{cases}
\end{equation*}
\begin{equation*} 
\bar{\theta}^1 := (\bar{\alpha}_1, \bar{\beta}_1), \quad \hbox{where}
\begin{cases}
\bar{\alpha}_1 \sim \text{Gamma}(\bar{a}_1,\bar{b}_1), \\
\bar{\beta}_1 \sim \text{N}(0, \sigma^2_{\bar{1}})
\end{cases}
\end{equation*}
\begin{equation*}
{\theta}^1 := ({\alpha}_1, {\beta}_1), \quad \hbox{where}
\begin{cases}
  \alpha_1 \sim \text{Gamma}(a_1,b_1), \\
  \beta_1 \sim \text{N}(0, \sigma^2_{1})
  \end{cases}
\end{equation*}
\begin{equation*}
\bar{\theta}^0 := (\bar{\alpha}_0, \bar{\beta}_0), \quad \hbox{where}
\begin{cases}
  \bar{\alpha}_0 \sim \text{Gamma}(\bar{a}_0,\bar{b}_0), \\
  \bar{\beta}_0 \sim \text{N}(0, \sigma^2_{\bar{0}})
  \end{cases}
\end{equation*}
\begin{equation*}
{\theta}^0 := ({\alpha}_0, {\beta}_0), \quad \hbox{where}
\begin{cases}
  \alpha_0 \sim \text{Gamma}(a_0,b_0), \\
  \beta_0 \sim \text{N}(0, \sigma^2_{0})
  \end{cases}
\end{equation*}
\begin{equation*}
    \bar{{\eta}}_{1} = {\eta}_{1} = \bar{{\eta}}_{0} = {\eta}_{0} \equiv {\eta}_Y \sim \text{N}({0},\Sigma_Y)
\end{equation*}
\begin{equation*}
    \delta \sim \text{N}(0, \sigma_{\delta}^2)
\end{equation*}
The results shown in the main text are obtained setting (i) all Gammas' shape parameters equal to $0.5$, (ii) all Gammas' scale parameters equal to $0.5$, (iii) all variances $\sigma^2 = 10^2$, and (iv) variance-covariance matrices $\Sigma$ as diagonal matrices with non-zero elements equal to $5^2$.
The results are quite robust to hyperparameters' specifications.

\section{Posterior distribution}\label{app3}
We defined the density of a Weibull random variable in Equation \ref{weibull}. 
Let the survival function of a generic Weibull distribution $Y$ be
\begin{equation}
    G_{Y}(y \mid a, b) = \exp\left\{-\left(\dfrac{y}{b}\right)^a\right\}
\end{equation}
Then, the density function of a generic left-truncated Weibull random variable is defined as:
\begin{equation}
    f_{Y}(y \mid a, b, l) = a y^{a-1} b^{-a} \exp\{b^{-a}(l^{a}-y^{a})\}\mathbb{I}_{l\leq y},
\end{equation}
where $l$ is the truncation value. 
Its survival will be
\begin{equation}
    G_{Y}(y \mid a, b, l) = \exp\left\{\left(\dfrac{l}{b}\right)^a-\left(\dfrac{y}{b}\right)\right\}\mathbb{I}_{l\leq y}\;.
\end{equation}

We write the posterior using the observed data likelihood:
\begin{equation} \label{obs_lik}
\begin{split}
    \pi({\theta}&\mid {X}, {Z},{C},\tilde{{D}},\tilde{{Y}})\propto \pi({\theta}^{\bar{\mathbb{D}}})\pi({\theta}^{D})\pi(\bar{{\theta}}^{1})\pi({\theta}^{1})\pi(\bar{{\theta}}^{0})\pi({\theta}^{0})\pi({\eta}_Y)\pi(\delta) \\
    & \times \prod\limits_{i:Z_i=1, \tilde{Y}_i = Y_i, \tilde{D}_i = C_i }{p({X}_i)f_{Y(1)}^{\bar{\mathbb{D}}}(Y_i|{X}_i,\bar{{\theta}}^{1},\eta_Y,\delta)} \\
    & \times \prod\limits_{i:Z_i=1,\tilde{Y}_i = C_i,\tilde{D}_i= D_i}{[1-p({X}_i)]f_{D(1)}(D_i|{X}_i,{\theta}^{D})G_{Y(1)}(C_i|D_i,{X}_i,{\theta}^{1},\eta_Y,\delta)} \\
    & \times \prod\limits_{i:Z_i=1,\tilde{Y}_i=Y_i,\tilde{D}_i=D_i}{[1-p({X}_i)]f_{D(1)}(D_i|{X}_i,{\theta}^{D})f_{Y(1)}(Y_i|D_i,{X}_i,{\theta}^{1},\eta_Y,\delta)} \\
    & \times \prod\limits_{i:Z_i=1,\tilde{Y}_i=C_i,\tilde{D}_i=C_i}{p({X}_i)G^{\bar{\mathbb{D}}}_{Y(1)}(C_i|{X}_i,\bar{{\theta}}^{1},\eta_Y) + [1-p({X}_i)]G_{D(1)}(C_i|{X}_i,{\theta}^{D})} \\
    & \times \prod\limits_{i:Z_i=0,\tilde{Y}_i = Y_i}\left[ p({X}_i)f_{Y(0)}^{\bar{\mathbb{D}}}(Y_i|{X}_i,\bar{{\theta}}^{0},\eta_Y) + 
     \right. \\ 
     & \left. 
    [1-p({X}_i)] \int\limits_{\mathbb{R}^{+}}f_{Y(0)}(Y_i|D_i(1) = d,{X}_i,{\theta}^{0},\eta_Y,\delta)f_{D(1)}(d|{X}_i,{\theta}^{D})~\mathrm{d}d  \right] \\
    & \times \prod\limits_{i:Z_i=0,\tilde{Y}_i = C_i}\left[ p({X}_i)G^{\bar{\mathbb{D}}}_{Y(0)}(C_i|{X}_i,\bar{{\theta}}^{0},\eta_Y) + 
    \right. \\ 
    & \left. 
    [1- p({X}_i)] \int\limits_{\mathbb{R}^{+}}G_{Y(0)}(C_i|D_i(1) = d, {X}_i,{\theta}^{0},\eta_Y,\delta)f_{D(1)}(d|{X}_i,{\theta}^{D})~\mathrm{d}d \right]
\end{split}
\end{equation} 
where the superscript $\bar{\mathbb{D}}$ denotes densities and survival functions for the ND stratum. 

Given the intractability of the observed data likelihood due to the presence of infinite mixtures, following Mattei et al. (2024), we include a data augmentation step simulating $D_i(1)$ for units assigned to the control arm.

We obtain the following complete data posterior distribution:
\begin{equation}
\begin{split}
    P({\theta}&|{X}, {Z},{C},\tilde{{D}},\tilde{{Y}},{D}(1),{Y}) \propto P({\theta})\mathcal{L}({\theta}|{X},{Z},{C},\tilde{{D}},\tilde{{Y}},{D}(1),{Y})\\[5pt]
    & \propto \pi({\theta^{\bar{\mathbb{D}}}})\pi({\theta}^{D})\pi(\bar{{\theta}}^{1})\pi({\theta}^{1})\pi(\bar{{\theta}}^{0})\pi({\theta}^{0})\\[5pt]
    & \times \prod_{i:Z_i=1,D_i(1) = \bar{\mathbb{D}}}{p({X}_i)f_{Y(1)}^{\bar{\mathbb{D}}}(Y_i|{X}_i,\bar{{\theta}}^{1},\eta_Y)^{\mathbb{I}\{Y_i\leq C_i\}}G_{Y(1)}^{\bar{\mathbb{D}}}(C_i|{X}_i,\bar{{\theta}}^{1},\eta_Y)^{\mathbb{I}\{Y_i> C_i\}}} \\[5pt]
    & \times \prod_{i:Z_i=1,D_i(1) \in \mathbb{R^{+}}}[1-p({X}_i)]G_{D(1)}(C_i|{X}_i,{\theta}^{D})^{\mathbb{I}\{D_i> C_i\}} \\[5pt]
    & \cdot \left[f_{D(1)}(D_i|{X}_i,{\theta}^{D})f_{Y(1)}(Y_i|D_i,{X}_i,{\theta}^{1},\eta_Y,\delta)^{\mathbb{I}\{Y_i\leq C_i\}} G_{Y(1)}(C_i|D_i,{X}_i,{\theta}^{1},\eta_Y,\delta)^{\mathbb{I}\{Y_i> C_i\}}\right]^{\mathbb{I}\{D_i\leq C_i\}} \\[5pt]
    & \times \prod_{i:Z_i=0,D_i(1) = \bar{\mathbb{D}}}{p({X}_i)f_{Y(0)}^{\bar{\mathbb{D}}}(Y_i|{X}_i,\bar{{\theta}}^{0},\eta_Y)^{\mathbb{I}\{Y_i\leq C_i\}}G_{Y(0)}^{\bar{\mathbb{D}}}(C_i|{X}_i,\bar{{\theta}}^{0},\eta_Y)^{\mathbb{I}\{Y_i> C_i\}}} \\[5pt]
    & \times \prod_{i:Z_i=0,D_i(1) \in \mathbb{R^{+}}}[1-p({X}_i)]f_{D(1)}(D_i(1)|{X}_i,{\theta}^{D}) 
    \\[5pt]
    & \cdot 
    f_{Y(0)}(Y_i|D_i(1),{X}_i,{\theta}^{0},\eta_Y,\delta)^{\mathbb{I}\{Y_i\leq C_i\}}G_{Y(0)}(C_i|D_i(1),{X}_i,{\theta}^{0},\eta_Y,\delta)^{\mathbb{I}\{Y_i> C_i\}}
\end{split}
\end{equation}

Once the full conditionals are derived, it will be possible to draw from the posterior above using a Metropolis-within-Gibbs algorithm such as the one described in Algorithm \ref{Alg1}.

\section{Algorithm description}\label{app5}
We implement an Adaptive Metropolis-within-Gibbs with Data Augmentation to estimate the causal effects of a new investigational drug, administered in addition to the standard of care, with respect to the standard of care only. 
The algorithm is described by Algorithms \ref{Alg1} and \ref{DA}.

\begin{algorithm}[!htbp]
\caption{Metropolis within Gibbs}\label{Alg1}
\SetAlgoLined
    Choose initial values for all the $J$ elements of the vector ${\theta}^{(0)}$ \; 
    Choose initial values $\text{I}^{\text{ND},(0)}_i ~\forall i\in \{Z_i = 0 \cup (Z_i = 1 \cap Y_i>C_i)\}$\; 
    Choose initial values $D_i(1)^{(0)} ~\forall i\in Z_i = 0$\;
    \For{$t\leftarrow 1$ \KwTo $T$}{
    \textbf{1. Data augmentation as in Algorithm \ref{DA}}\\
    \textbf{2. Parameters' update}\\
    \For{${\theta}_j$ \KwTo ${\theta}_1$:${\theta}_J$}{
    draw ${\theta}_j^*$ from a proposal distribution $q_t({\theta}_j^*|{\theta}_j^{(t-1)})$ \;
    compute the acceptance ratio 
    $\nu_{{\theta}_j} = \min\left(1; \frac{\pi({\theta}_j^*\mid \cdot)}{\pi({\theta}_j^{(t-1)}\mid \cdot)}\frac{q_t({\theta}_j^{(t-1)}|{\theta}_j^*)}{q_t({\theta}_j^*|{\theta}_j)}\right)$ \;
    draw $u \sim \text{Unif}(0,1)$ \;
    \eIf{$u<\nu_{{\theta}_j}$}{
    set ${\theta}_j^{(t)} = {\theta}_j^*$}
    {set ${\theta}_j^{(t)} = {\theta}_j^{(t-1)}$}}}
\end{algorithm}

\begin{algorithm}[!htbp]
\caption{Data augmentation step at iteration $t$}\label{DA}
\SetAlgoLined
    \For{$i: ~Z_i = 1, Y_i>C_i$}{
    compute \\ 
    $\zeta_i = \frac{p({X}_i, {\theta}^{\bar{\mathbb{D}},(t)})G_{Y(1)}^{\bar{\mathbb{D}}}(Y_i\mid {X}_i, {\theta}^{1, (t)})}{p({X}_i, {\theta}^{\bar{\mathbb{D}},(t)})G_{Y(1)}^{\bar{\mathbb{D}}}(Y_i\mid {X}_i, {\theta}^{1,(t)})+(1-p({X}_i, {\theta}^{\bar{\mathbb{D}},(t)}))G_{D(1)}(D_i^\text{obs}\mid {X}_i, {\theta}^{1,(t)})}$ \; 
    draw I$^{\text{ND}*}_i \sim \text{Ber}(\zeta_i)$}
    \For{$i: ~Z_i = 0$}{
    draw I$^{\text{ND}*}_i \sim \text{Ber}(p({X}_i, {\theta}^{\bar{\mathbb{D}},(t-1)}))$ \;
    \eIf{$\text{I}^{\text{ND}*}_i = 0$}
    {draw $D(1)_i^{*} \sim \text{Weibull}(\alpha_D^{(t-1)},\exp{(-\{\beta_D^{(t-1)}+{X}_i'{\eta}_D^{(t-1)}\}/\alpha_D^{(t-1)})})$}
    {set $D(1)_i^{*} = 0$}
    compute the ratio $r_i = \frac{\pi({\theta}^{(t-1)}\mid \text{I}^{\text{ND}*}_i, D_i^*, Y_i, {X}_i)}{\pi({\theta}^{(t-1)}\mid \text{I}^{\text{ND},(t-1)}_i, D_i^{(t-1)}, Y_i, {X}_i)}$ \;
    compute $\rho_i = \begin{cases}
        1 & \text{if~I}^{\text{ND}*}_i = \text{I}^{\text{ND},{(t-1)}}_i = 1  \\[10pt]
        \frac{p({X}_i,{\theta}^{\bar{\mathbb{D}},(t-1)})}{(1-p({X}_i,{\theta}^{\bar{\mathbb{D}},(t-1)}))f_D(D_i^{*}\mid {X}_i, {\theta}^{D,(t-1)})}& \text{if~I}^{\text{ND}*}_i = 0, \text{I}^{\text{ND},{(t-1)}}_i = 1 \\[10pt]
        \frac{(1-p({X}_i,{\theta}^{\bar{\mathbb{D}},(t-1)}))f_D(D_i^{(t-1)}\mid {X}_i, {\theta}^{D,(t-1)})}{p({X}_i,{\theta}^{\bar{\mathbb{D}},(t-1)})} & \text{if~I}^{\text{ND}*}_i = 1, \text{I}^{\text{ND},{(t-1)}}_i = 0  \\[10pt]
        \frac{f_D(D_i^{(t-1)}\mid {X}_i, {\theta}^{D,(t-1)})}{f_D(D_i^{*}\mid {X}_i, {\theta}^{D,(t-1)})}  & \text{if~I}^{\text{ND}*}_i = \text{I}^{\text{ND},{(t-1)}}_i = 0 
    \end{cases}$ \;
    \eIf{$u<(r_i \cdot \rho_i)$}
    {set $\text{I}^{\text{ND},{(t)}}_i = \text{I}^{\text{ND}*}_i$ and $D(1)_i^{(t)} = D(1)_i^{*}$}
    {set $\text{I}^{\text{ND},(t)}_i = \text{I}^{\text{ND},(t-1)}_i$ and $D(1)_i^{(t)} = D(1)_i^{(t-1)}$}
    }
\end{algorithm}

\newpage
\section{Summary statistics for the two scenarios}\label{app6}
\begin{table}[h]
\centering
\caption{Summary data, scenario I}
{\begin{tabular}{@{}llcccccc@{}}
\hline
\textbf{Variable}    & \textbf{Mean(proportion)} & \textbf{SD} & \textbf{Min} & \textbf{Q1} & \textbf{Median} 
& \textbf{Q3} & \textbf{Max} \\ \hline
$Z_i$                & 54.0\% (181/335)          & -----       & -----        & -----       & ----- 
& -----       & -----  \\ 
$\mathbb{I}(D_i^\text{obs}<C_i)$   & 14.93\%(50/335)  & -----       & -----        & -----  & -----        
& ----- & ----- \\
$D^\text{obs}_i$     & 2.99  (50/335)              & 2.41         & 0.19       & 1.04   & 2.64 & 
3.87 & 11.99\\ 
$\mathbb{I}(Y_i^\text{obs}<C_i)$ & 91.64\% (307/335)    & ----- & ----- & ----- & ----- 
& ----- & -----\\
$Y^\text{obs}$        & 5.16 (307/335) & 3.66 & 0.02 & 2.31 & 4.35 
& 7.58 & 20.76 \\  \textbf{Continuous covariates}                  &                  &        &       &         \\ 
 $X_1$                                     & 63.09            & 10.46 & 28.44  & 55.80 & 63.09 & 70.48 & 91.51  \\ 
 \textbf{Categorical covariates}    &      &  &  &  &  
&  & \\
 $X_2$ & 43.88\% (147/335)    & ----- & ----- & ----- & ----- 
& ----- & -----\\
$X_3$ & 25.37\% (85/335)    & ----- & ----- & ----- & ----- 
& ----- & -----\\
\hline
\end{tabular}}
\end{table}

\begin{table}[h]
\centering
\caption{Summary data, scenario II}
{\begin{tabular}{@{}llcccccc@{}}
\hline
\textbf{Variable}    & \textbf{Mean(proportion)} & \textbf{SD} & \textbf{Min} & \textbf{Q1} & \textbf{Median} 
& \textbf{Q3} & \textbf{Max} \\ \hline
$Z_i$                & 54.0\% (181/335)          & -----       & -----        & -----       & ----- 
& -----       & -----  \\ 
$\mathbb{I}(D_i^\text{obs}<C_i)$   & 14.93\%(50/335)  & -----       & -----        & -----  & -----        
& ----- & ----- \\
$D^\text{obs}_i$     & 2.99  (50/335)              & 2.41         & 0.19       & 1.04   & 2.64 & 
3.87 & 11.99\\ 
$\mathbb{I}(Y_i^\text{obs}<C_i)$ & 90.15\% (302/335)    & ----- & ----- & ----- & ----- 
& ----- & -----\\
$Y^\text{obs}$        & 5.51 (302/335) & 3.87 & 0.01 & 2.36 & 4.87 
& 7.84 & 20.76 \\  \textbf{Continuous covariates}                  &                  &        &       &         \\ 
 $X_1$                                     & 63.09            & 10.46 & 28.44  & 55.80 & 63.09 & 70.48 & 91.51  \\ 
 \textbf{Categorical covariates}    &      &  &  &  &  
&  & \\
 $X_2$ & 43.88\% (147/335)    & ----- & ----- & ----- & ----- 
& ----- & -----\\
$X_3$ & 25.37\% (85/335)    & ----- & ----- & ----- & ----- 
& ----- & -----\\
\hline
\end{tabular}}
\end{table}

\end{document}